\documentclass[twocolumn,showpacs,pra,aps,graphicx,preprintnumbers,amsmath,amssymb]{revtex4}

\usepackage{dcolumn}
\usepackage{bm}

\usepackage{epic}
\usepackage{eepic}
\usepackage{graphicx}

\newcommand{\ket}[1]{|#1\rangle}

\renewcommand{\vec}[1]{{\bf{#1}}}

\newcommand{\Order}[1]{\mathcal{O}(#1)}

\begin{document}

\title{Molecule formation as a diagnostic tool for second order correlations
of ultra-cold gases}
\author{D. Meiser}
\author{P. Meystre}
\affiliation{Optical Sciences Center, The University of Arizona,
Tucson, Arizona 85721}
\author{C. P. Search}
\affiliation{Department of Physics and Engineering Physics, Stevens Institute of Technology, Hoboken, New Jersey 07030}

\pacs{39.20.+q,03.75.-b,74.90.+n}

\begin{abstract}
We calculate the momentum distribution and the second-order
correlation function in momentum space, $g^{(2)}({\bf p},{\bf
p}',t)$ for molecular dimers that are coherently formed from an
ultracold atomic gas by photoassociation or a Feshbach resonance.
We investigate using perturbation theory how the quantum
statistics of the molecules depend on the initial state of the
atoms by considering three different initial states: a
Bose-Einstein condensate (BEC), a normal Fermi gas of ultra-cold
atoms, and a BCS-type superfluid Fermi gas. The cases of strong
and weak coupling to the molecular field are discussed. It is
found that BEC and BCS states give rise to an essentially coherent
molecular field with a momentum distribution determined by the
zero-point motion in the confining potential. On the other hand, a
normal Fermi gas and the unpaired atoms in the BCS state give rise
to a molecular field with a broad momentum distribution and
thermal number statistics. It is shown that the first-order
correlations of the molecules can be used to measure second-order
correlations of the initial atomic state.
\end{abstract}

\maketitle

\section{\label{introduction} Introduction}

The basis of some of the most exciting developments in ultra-cold
atomic physics in recent years has been the use of Feshbach
resonances \cite{Inouye:MoleculeFR,Timmermans:Feshbachresonances}
and photoassociation \cite{Fedichev:,Theis:} to tune the strength
of the interactions between atoms as wells as to create ultracold
diatomic molecules starting from an ultracold atomic gas of bosons
or fermions
\cite{Wynar:MoleculePa,Inouye:MoleculeFR,Donley:MoleculeBEC,Duerr:MoleculeRP,Greiner:MoleculeBEC,Zwierlein:Li2,Jochim:Li2}.
The availability of tunable interactions has made possible the
study of many model systems of condensed matter theory in a very
controlled fashion
\cite{Bloch:MottInsulator1,Bloch:MottInsulator2,Jaksch:BECinLattice,Stoof:Varenna1998,Timmermans:Fermigasphysics,Timmermans:BCS}.
In particular, the BEC-BCS crossover, which predicts a continuous
transition from a BCS superfluid of atomic fermions to a molecular
BEC as the interaction strength is varied from attractive to
repulsive, has attracted a considerable amount of attention, both
experimentally and theoretically
\cite{Timmermans:BCS,Regal:BEC_BCScrossover,Bartenstein:BEC_BCScrossover,Zwierlein:BEC_BCScrossover,Stoof:Li6,Bruun:BCS_theory}.

A difficulty in the studies of the BEC-BCS crossover has been that
they necessitate the measurement of higher order correlations of
the atomic system. While the momentum distribution of a gas of
bosons provides a clear signature of the presence of a
Bose-Einstein condensate, the Cooper pairing between fermionic
atoms in a BCS state hardly changes the momentum distribution or
spatial profile as compared to a normal Fermi gas. This poses a
significant experimental challenge, since the primary techniques
for probing the state of an ultracold gas are either optical
absorption or phase contrast imaging, which directly measure the
spatial density or momentum distribution following ballistic
expansion of the gas. In the strongly interacting regime very
close to the Feshbach resonance, evidence for fermionic
superfluidity was obtained by projecting the atom pairs onto a
molecular state by a rapid sweep through the resonance
\cite{Regal:BEC_BCScrossover,Zwierlein:BEC_BCScrossover}. More
direct evidence of the gap in the excitation spectra due to
pairing was obtained by rf spectroscopy \cite{Chin:} and by
measurements of the collective excitation frequencies
\cite{Kinast:,Bartenstein:}.

Still, the detection of fermionic superfluidity in the weakly
interacting BCS regime remains a challenge. The direct detection
of Cooper pairing requires the measurement of second-order or
higher atomic correlation functions. Several researchers have
proposed and implemented schemes that allow one to measure higher
order correlations
\cite{Burt:BEC_decay,Hellweg:Spatial_correlations,Cacciapuoti:Second_correlation_function,Altman:noise_correlations,Regal:BEC_BCScrossover,Radka:Diagnosing_correlations}
but those methods are still very difficult to realize
experimentally.

While the measurement of higher order correlations is challenging
already for bosons, the theory of these correlations has been
established a long time ago by Glauber for photons
\cite{Glauber:Optical_Coherence1,Glauber:Optical_Coherence2,Naraschewski:Spatial_coherence}.
For fermions however, despite some efforts
\cite{Cahill:Density_Operator_Fermions} a satisfactory coherence
\emph{theory} is still missing.

To circumvent these difficulties we suggested in an earlier
publication \cite{Meiser:singlemode_molecules}, guided by the
analogy with three-wave mixing in classical optics, to make use of
the nonlinear coupling of atoms to a molecular field by means of a
two-photon Raman transition or a Feshbach resonance. The
nonlinearity of the coupling links first-order correlations of the
molecules to second-order correlations of the atoms. Furthermore
the molecules are always bosonic so that the well-known coherence
theory for bosonic fields can be used to characterize them.
Considering a simplified model with only one molecular mode, it
was found that the molecules created that way can indeed be used
as a diagnostic tool for second-order correlations of the original
atomic field. Naturally, due to the restriction to a single mode,
the information one can gain about the atomic state is very
limited.

In this paper we extend the previous model to take into account
all modes of the molecular field, the hope being that in doing so,
more detailed information about the atomic state can be obtained.
Specifically, we calculate the momentum distribution of the
molecules and the normalized second-order correlation function,
$g^{(2)}$ for different momentum states of the molecules using
perturbation theory. We consider the limiting cases of strong or
weak atom-molecule coupling as compared to the relevant atomic
energies. The molecule formation from a Bose-Einstein condensate
(BEC) serves as a reference system. There we can rather easily
study the contributions to the molecular signal from the condensed
fraction as well as from thermal and quantum fluctuations above
the condensate. The cases of a normal Fermi gas and a BCS
superfluid Fermi system are then compared with it. We show that
the molecule formation from a normal Fermi gas and from the
unpaired fraction of atoms in a BCS state has very similar
properties to those of the molecules formed from the non-condensed
atoms in the BEC case. The state of the molecular field formed
from the pairing field in the BCS state on the other hand is
similar to that resulting from the condensed fraction in the BEC
case. The qualitative information gained by the analogies with the
BEC case help us gain a physical understanding of the molecule
formation in the BCS case where direct calculations are difficult
and not nearly as transparent.

This paper is organized as follows: In section \ref{model} we
introduce the model Hamiltonian used to describe the coupled
atom-molecule system. In sections \ref{BEC} to \ref{BCS} we
present the calculations of momentum distribution and second
factorial moment of the molecular field for a BEC, a normal Fermi
gas and a BCS-type state, respectively, considering the cases of
strong and weak coupling in each case. Details of the calculations
are given in the appendices \ref{appendixa} and \ref{appendixb}.

\section{\label{model} Model}

The general procedure that we have in mind is the following: The
atomic sample is prepared in some initial state, whose higher
order correlations we seek to analyze. At time $t=0$ the coupling
to a molecular field is switched on. While the initial atomic
state corresponds to a trapped gas, we assume that the molecules
can be treated as free particles. This is justified if the atomic
trapping potential does not affect the molecules, or if the
interaction time between the atoms and molecules is much less than the
oscillation period in the trap. Finally, the state of the molecular field is
analyzed by standard techniques, e.g. time of flight measurements.

We consider the three cases where the atoms are bosonic and initially form a
BEC, or consist of two species of ultra-cold fermions (labeled by
$\sigma=\uparrow,\downarrow$), with or without superfluid component. In the
following we describe explicitly the situation for fermions, the bosonic case
being obtained from it by omitting the spin indices and by replacing the Fermi
field operators by bosonic field operators.

Since we are primarily interested in how much can be learned about
the second-order correlations of the initial atomic cloud from the
final molecular state, we keep the physics of the atoms themselves
as well as the coupling to the molecular field as simple as
possible. The coupled fermion-molecule system can be described by
the Hamiltonian
\cite{Holland:BCS,Chiofalo:Res_superfluidity,Timmermans:Feshbachresonances}
\begin{widetext}
\begin{eqnarray}
\hat{H}&=&\sum_{\vec{k},\sigma}
\epsilon_\vec{k}\hat{c}_{\vec{k}\sigma}^\dagger
\hat{c}_{\vec{k}\sigma}+
\sum_\vec{k}E_\vec{k}\hat{a}_{\vec{k}}^\dagger \hat{a}_{\vec{k}}+
V^{-1/2}\sum_{\vec{k}_1,\vec{k}_2,\sigma}
\tilde{U}_\text{tr}(\vec{k}_2-\vec{k}_1)
\hat{c}_{\vec{k}_2\sigma}^\dagger
\hat{c}_{\vec{k}_1\sigma}\nonumber \\
&+& \frac{U_0}{2V}\sum_{\vec{q},\vec{k}_1,\vec{k}_2}
\hat{c}_{\vec{k}_1+\vec{q}\uparrow}^\dagger
\hat{c}_{\vec{k}_2-\vec{q}\downarrow}^\dagger
\hat{c}_{\vec{k_2}\downarrow} \hat{c}_{\vec{k}_1\uparrow}+
g\left(\sum_{\vec{q},\vec{k}}\hat{a}_{\vec{q}}^\dagger
\hat{c}_{\vec{q}/2+\vec{k}\downarrow}\hat{c}_{\vec{q}/2-\vec{k}\uparrow}
+ \text{H.c.}\right) \label{full_hamiltonian}
\end{eqnarray}
\end{widetext}
Here $\epsilon_\vec{k}=k^2/2M$ is the kinetic energy of an atom of mass
$M$ and momentum $\vec{k}$ and $E_\vec{k}=\epsilon_\vec{k}/2 +\nu$
is the energy of a molecule with momentum $k$ and detuning parameter $\nu$.
$c_{\vec{k}\sigma}$ and $c_{\vec{k}\sigma}^\dagger$ are fermionic annihilation
and creation operators for plain waves in quantization volume $V$ with
spin $\sigma$.
$\tilde{U}_\text{tr}(\vec{k})=V^{-1/2}\int_V d^3x
e^{-i\vec{k}\vec{x}} U_{\rm tr}(\vec{x})$ is the Fourier transform
of the trapping potential $U_{\rm tr}({\bf r})$ and $U_0 = 4\pi a/M$ is the
background scattering strength, $g$ is the effective coupling constant of the
atoms to the molecules and we use units with $\hbar\equiv 1$
throughout. We assume that the trapping potential and background
scattering are relevant only for the preparation of the initial
state before the coupling to the molecules is switched on at $t=0$
and can be neglected in the calculation of the dynamics. This is
justified if $g\sqrt{N} \gg U_0n,\omega_i$ where $n$ is the
atomic density, $N$ the number of atoms, and $\omega_i$ are the
frequencies of $U_{\rm tr}({\bf r})$ that is assumed to be harmonic.
In experiments, the interaction between the atoms can effectively be
switched off by ramping the magnetic field to a position where the
scattering length is zero.

Regarding the strength of the coupling constant $g$, two cases are
possible: $g\sqrt{N}$ can be much larger or much smaller than the
characteristic kinetic energies involved. For fermions the terms broad and
narrow resonance have been coined for the two cases,
respectively, and we will use these for bosons as well. Both
situations can be realized experimentally, and they give rise to
different effects. We examine both limiting cases and 
use the suggestive notation $E_{\rm kin}\ll g\sqrt{N}$ and
$E_{\rm kin}\gg g\sqrt{N}$ for the two cases, where $E_{\rm kin}$ denotes
the characteristic kinetic energy of the atoms. It
corresponds to zero point motion for condensate atoms, to the
thermal energy, $k_BT$ for non-condensed thermal bosons, and to
the Fermi energy for a degenerate Fermi gas.

Our analysis is based on the assumption that first order time-dependent 
perturbation theory is applicable. This requires that
the state of the atoms does not change significantly and consequently,
only a small fraction of the atoms are converted into molecules.
It is reasonable to assume that this is true for short interaction
times or weak enough coupling. Apart from making the system
tractable by analytic methods there is also a deeper reason 
why the coupling should be weak: Since we ultimately wish to get
information about the atomic state, it should not be modified too
much by the measurement itself, i.e. the coupling to the molecular
field. Our treatment therefore follows the same spirit as Glauber's
original theory of photon detection, where it is assumed that the
light-matter coupling is weak enough that the detector
photocurrent can be calculated using Fermi's Golden rule.

\section{\label{BEC} BEC}

We consider first the case where the initial atomic state is a BEC
in a spherically symmetric harmonic trap. We note that all of
our results can readily be extended to anisotropic traps by an
appropriate rescaling of the coordinates in the direction of the
trap axes. We assume that the temperature is well below the
BEC transition temperature and that the interactions between the atoms
are not too strong. Then the atomic system is described by the
field operator
\begin{equation}
\hat{\psi}(\vec{x})=\chi_0(\vec{x})\hat{c}+\delta\hat{\psi}(\vec{x}),
\label{decomposition_psi}
\end{equation}
where
\begin{equation}
\chi_0(\vec{x})=\sqrt{\frac{15}{8\pi R_{\rm
TF}^3}}\sqrt{1-\frac{x^2}{R_{\rm TF}^2}}
\label{TFwavefunction}
\end{equation}
is the condensate wave function in the Thomas-Fermi approximation
and $\hat{c}$ is the annihilation operator for an atom in the
condensate, $R_{\rm TF}=(15Na/a_{\rm osc})^{1/5}a_{\rm osc}$ is
the Thomas-Fermi radius, $N$ the number of atoms, $a$ is their
scattering length and $a_{\rm osc}$ is the oscillator length of
the atoms in the trap. In accordance with the assumption of low
temperatures and weak interactions we do not distinguish
between the total number of atoms and the number of atoms in the
condensate. The fluctuations $\delta \hat{\psi}(\vec{x})$ are small
and those with wavelengths much less than $R_{TF}$ will be treated
in the local density approximation while those with wavelengths
comparable to $R_{TF}$ can be neglected.
\cite{Hutchinson:Finite_T_BEC,Reidl:Finite_T_BEC,Bergeman:BEC_Tc}.

\subsection{Broad resonance, $E_{\rm kin}\ll g\sqrt{N}$}

We are interested in the momentum distribution of the molecules
\begin{equation}
n(\vec{p},t)=\langle \hat{a}_\vec{p}^\dagger(t) \hat{a}_\vec{p}(t)\rangle
\end{equation}
which for short times, $t$, can be calculated using perturbation theory: We
expand $n(\vec{p},t)$ in a Taylor series around $t=0$ and make use use of the
Heisenberg equations of motion
\begin{equation}
i\frac{\partial \hat{a}_\vec{p} (t)}{\partial t}=
g\sum_\vec{k}\hat{c}_{\vec{p}/2+\vec{k}}\hat{c}_{\vec{p}/2-\vec{k}},
\label{eqnofmotion_BEC}
\end{equation}
and similarly for $\hat{a}^\dagger_{\vec{p}}$. Here we have
neglected the kinetic energy term, a step that is legitimate for a broad
resonance since the interaction energy $g\sqrt{N}$ is much larger
than the difference in the kinetic energies between the atoms and
molecules. Consequently for short enough times, $t\lesssim
(g\sqrt{N})^{-1}$, energy conservation can be violated in the
formation of molecules in a fashion similar to the Raman-Nath regime
of atomic diffraction.

To lowest non-vanishing order in $gt$ we find
\begin{widetext}
\begin{equation}
n_{\rm BEC,b}(\vec{p},t)=(gt)^2\sum_{\vec{k_1},\vec{k_2}}\left\langle
\hat{c}_{\vec{p}/2-\vec{k_1}}^\dagger \hat{c}_{\vec{p}/2+\vec{k_1}}^\dagger
\hat{c}_{\vec{p}/2+\vec{k_2}}\hat{c}_{\vec{p}/2-\vec{k_2}}\right\rangle
+\Order{(gt)^4},
\label{g1_BEC_explicit}
\end{equation}
\end{widetext}
where the atomic operators are the initial ($t=0$) operators. This
expression can be evaluated by making use of the decomposition of
the atomic field operator \eqref{decomposition_psi} and the local
density approximation for the part describing the fluctuations.
The details of this calculation are given in appendix
\ref{appendixa}. To first non-vanishing order in the fluctuations
we find
\begin{widetext}
\begin{equation}
\label{g1bec}
n_{\rm BEC,b}(\vec{p},t)=
(gt)^2 N(N-1)V\left|{\tilde{\chi}_0^2}(\vec{p})\right|^2
 +(gt)^2 4N\int \frac{d^3x}{V}
 \left\langle \delta\hat{c}_\vec{p}^\dagger(\vec{x}) \delta\hat{c}_\vec{p}(\vec{x})\right\rangle.
\end{equation}
\end{widetext}
From this expression we see that our approach is justified if
$(\sqrt{N}gt)^2\ll 1$ because for such times the initial atomic
state can be assumed to remain undepleted. 

In the local density approximation the expectation value $\left\langle
\delta\hat{c}_\vec{p}^\dagger(\vec{x})
\delta\hat{c}_\vec{p}(\vec{x})\right\rangle$ for the number of
fluctuations with momentum $\vec{p}$ at $\vec{x}$ can be
evaluated by assuming that at each $\vec{x}$ we have a homogenous
BEC with density $n(\vec{x})$ and using the Bogoliubov
transformation to quasi-particle operators,
$\hat{\alpha}_{\vec{k}}(\vec{x})$. One finds 
\begin{equation}
\left\langle \delta\hat{c}_\vec{p}^\dagger(\vec{x})
\delta\hat{c}_\vec{p}(\vec{x})\right\rangle=
v_\vec{p}^2(\vec{x}) +
(u_\vec{p}^2(\vec{x})+v_\vec{p}^2(\vec{x}))\left\langle {\hat{\alpha}^\dagger_\vec{p}(\vec{x})}
\hat{\alpha}_\vec{p}(\vec{x})\right\rangle,
\end{equation}
with Bogoliubov amplitudes
\begin{gather}
u_\vec{p}^2(\vec{x})=\frac{1}{2}\left[\frac{\frac{p^2}{2M}+n(\vec{x})U_0}
{\tilde{\epsilon}_\vec{p}(\vec{x})}+1\right],\\
v_\vec{p}^2(\vec{x})=\frac{1}{2}\left[\frac{\frac{p^2}{2M}+n(\vec{x})U_0}
{\tilde{\epsilon}_\vec{p}(\vec{x})}-1\right],
\end{gather}
and quasi-particle energies
\begin{equation}
\tilde{\epsilon}_\vec{p}(\vec{x})=\sqrt{\epsilon_\vec{p}^2+2\epsilon_{\vec{p}}n(\vec{x})U_0}.
\end{equation}
The quasi-particle distribution is given by a thermal Bose distribution
\begin{equation}
\left\langle
\hat{\alpha}_\vec{p}^{\dagger}(\vec{x})\hat{\alpha}_\vec{p}(\vec{x})\right\rangle =
\frac{1}{e^{\tilde{\epsilon}_\vec{p}(\vec{x})/k_BT}-1}.
\end{equation}

\begin{figure}
\includegraphics[width=0.9\columnwidth]{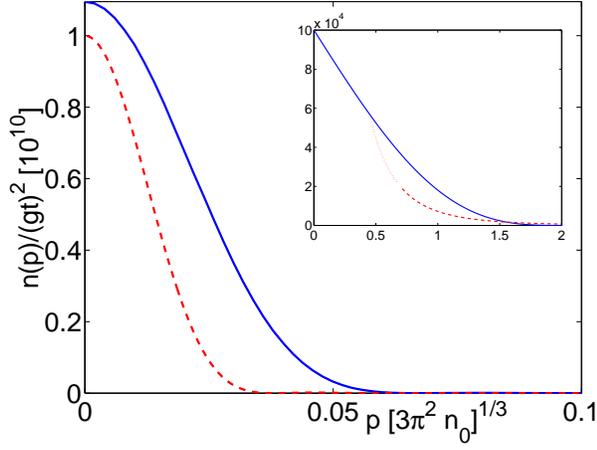}
\caption{(color online) Momentum distribution of molecules formed from a BEC
(red dashed line) with $a=0.1a_{\rm osc}$ and $T=0.1T_c$ and a BCS type state
with $k_F a=0.5$ and $a_{\rm osc}=5k_F^{-1}(0)$ (blue solid line), both for
$N=10^5$ atoms. The BCS curve has been scaled up by a factor of $20$ for easier
comparison. The inset shows the noise contribution for BEC (red dashed) and BCS
(blue) case.  The latter is simply the momentum distribution of molecules
formed from a normal Fermi gas. The local density approximation treatment of
the noise contribution in the BEC case is not valid for momenta smaller than
$2\pi/\xi$ (indicated by the red dotted line in the inset). Note that the
coherent contribution is larger than the noise contribution by five orders of
magnitude in the BEC case and three orders of magnitude in the BCS case.}
\label{nofp_broad}
\end{figure}

The momentum distribution \eqref{g1bec} is illustrated in Fig.
\ref{nofp_broad}. The contribution from the condensate is a
collective effect, as indicated by its quadratic scaling with the
atom number. It clearly dominates over the incoherent contribution
from the fluctuations, which is proportional to the number of
atoms. The momentum width of the contribution from
the condensate is roughly $2\pi/R_{\rm TF}$ which is much narrower
than the contribution from the fluctuations, whose momentum
distribution has a typical width of $1/\xi$, where $\xi=(8\pi a
n)^{-1/2}$ is the healing length. Using the Thomas-Fermi wave
function for the condensate, Eq. \eqref{TFwavefunction}, we can
calculate the condensate contribution in closed form as
\begin{widetext}
\begin{equation}
n_{\rm BEC,b}(\vec{p},t)
=\frac{225 N(N-1)(gt)^2}{4(pR_{TF})^6}
\left(\frac{6\sin pR_{TF}}{(pR_{TF})^2} - \frac{6\cos pR_{TF}}{pR_{TF}}
-2\sin pR_{TF}\right)^2 + \Order{N}.
\label{nBEC_TF}
\end{equation}
\end{widetext}
The terms of order $N$ are corrections due to the non-condensed part.

Using the same approximation scheme we can calculate the second-order correlation
\begin{equation}
g^{(2)}(\vec{p}_1,t_1;\vec{p}_2,t_2)= \frac{\langle
\hat{a}_{\vec{p}_1}^\dagger (t_1)\hat{a}_{\vec{p}_2}^\dagger
(t_2)\hat{a}_{\vec{p}_2}(t_2) \hat{a}_{\vec{p}_1}(t_1)}
{n({\vec{p}_1},t_1)n({\vec{p}_2},t_2)}. \label{general_g2}
\end{equation}
If we neglect fluctuations we find
\begin{eqnarray}
g_{\rm BEC,b}^{(2)}(\vec{p}_1, t_1; \vec{p}_2, t_2)&=&
\frac{(N-2)!^2}{(N-4)!N!}\nonumber\\
&=&1-\frac{6}{N}+\Order{N^{-2}}.
\label{g2perturbative_bec}
\end{eqnarray}
For $N\rightarrow \infty$ this is very close to 1, which is
characteristic of a coherent state. This
result implies that the number fluctuations of the molecules are
very nearly Poissonian. The fluctuations lead to a larger value of
$g^{(2)}$, making the molecular field partially coherent, but
their effect is only of order $\Order{N^{-1}}$.

The physical reason why the resulting molecular field is almost
coherent is of course clear: The condensed fraction of the atomic
field operator is dominant. In expectation values the operators
$\hat{c}$ and $\hat{c}^\dagger$ take on values $\sqrt{N-n}$, with a
number $n\ll N$ depending on the position of the operator in the
expectation value. When dividing by the normalizing expectation
values $n(\vec{p},t)$, $\sqrt{N-n}$ can be replaced by $\sqrt{N}$
with accuracy $\Order{N^{-1}}$ and hence $\hat{c}$ and
$\hat{c}^\dagger$ can be replaced by $\sqrt{N}$ independent of
their position in the expectation value. The field operator can
thus be replaced by a $c$-number field $\hat{\psi}(x)\to
\sqrt{N}\chi_0(x)$, the mean field, which explains the almost
perfect factorization of the correlation functions.

Another way to understand this is to consider the single-mode BEC
state,
$|\Psi(0)\rangle=(\hat{c}^{\dagger}\hat{c}^{\dagger})^{(N/2)}|0\rangle/\sqrt{N!}$.
The coupling to the molecular field will transform this state into
$|\Psi(t)\rangle\approx(\alpha\hat{a}^{\dagger}+\beta\hat{c}^{\dagger}\hat{c}^{\dagger})^{(N/2)}|0\rangle/\sqrt{N!}$,
where $\alpha \ll 1$. This leads to a binomial distribution for
the number of molecules. In the limit that $N\rightarrow \infty$
the Binomial distribution goes over to the Poisson distribution.

\subsection{Narrow resonance, $E_{\rm kin}\gg g\sqrt{N}$}

For a narrow resonance, the typical kinetic energies associated
with the atoms and molecules, $E_{\rm kin}$, are much larger than the
atom-molecule interaction energy. This implies that even for very
short interaction times, $t\lesssim (g\sqrt{N})^{-1}$, the phase
of the atoms and molecules can evolve significantly, $E_{\rm kin} t\gg
1$. Consequently, only transitions between atom pairs and
molecules that conserve energy can occur. In this case, it is
convenient to go over to the interaction representation
\begin{equation}
\hat{c}_{\bf{p}}(t)\rightarrow
e^{-i\epsilon_{\bf{p}}t}\hat{c}_{\bf{p}}(t),\quad
\hat{a}_{\bf{p}}(t)\rightarrow e^{-iE_{\bf{p}}t} \hat{a}_{\bf
{p}}(t),
\end{equation}
where for notational convenience, we will denote the interaction picture operators
by the same symbols as the Heisenberg operators used in the
previous subsection.

The equations of motion in the interaction picture,
\begin{equation}
i\frac{\partial \hat{a}_\vec{p} (t)}{\partial t}
=g \sum_\vec{k}
e^{i\left(E_\vec{p}-\epsilon_{\vec{p}/2+\vec{k}}-\epsilon_{\vec{p}/2-\vec{k}}\right)t}
\hat{c}_{\vec{p}/2+\vec{k}}\hat{c}_{\vec{p}/2-\vec{k}},
\label{eqnofmotion_BEC_lt}
\end{equation}
can be approximately integrated by treating the atomic operators
as constants, leading to
\begin{equation}
\hat{a}_\vec{p}(t)=\sum_{\vec{k}}\Delta(E_p-\epsilon_{\vec{p}/2+\vec{k}}-\epsilon_{\vec{p}/2-\vec{k}},t)\hat{c}_{\vec{p}/2+\vec{k}}\hat{c}_{\vec{p}/2-\vec{k}},
\label{aofp_lt}
\end{equation}
where we have introduced
\begin{equation}
\Delta(\omega,t)=g\lim_{\eta\downarrow 0}\frac{e^{i\omega
t}-1}{i\omega+\eta}.
\label{DefDelta}
\end{equation}
\footnote{We denote the function in Eq. \eqref{DefDelta} by $\Delta$ because it has properties similar to the usual $\delta$-function. Confusion with the local gap parameter introduced below, which we also denote by $\Delta$, cannot arise because the first always has energies as its argument while the latter has positions as its argument.}

The condition under which this step is justified is analyzed below.
As in the broad resonance case we can insert this expression in
$n(\vec{p},t)$. The calculation of the resulting integrals over expectation
values of the atomic state is however considerably subtler than in for the broad resonance case and is presented in details in
appendix \ref{appendixb}. In the limit $\nu t\gg 1$ we find
\begin{widetext}
\begin{eqnarray}
n_{\rm BEC,n}(\vec{p},t)&=&N(N-1)\frac{V^2M^3 g^2}{16\pi^4}
\left|\int_0^\infty d\omega \sqrt{\omega}
\left(\pi\delta(\nu-\omega)+i{\bf P}\frac{1}{\nu-\omega}\right)\right.\nonumber\\
&\times &\left.\int_{-1}^{1} dz \tilde{\chi}_0\left(\sqrt{p^2/4 + M\omega - p\sqrt{M\omega}z}\right) \tilde{\chi}_0\left(\sqrt{p^2/4+M\omega+p\sqrt{M\omega}z}\right)\right|^2\nonumber\\
&+&N\delta_{p/2,\sqrt{M\nu}}\frac{3g^2 t \sqrt{\nu M^3}R_{TF}^3}{8\pi^2}\int \frac{d^3x}{V}\langle \delta\hat{c}_\vec{p}^\dagger(\vec{x}) \delta\hat{c}_\vec{p}(\vec{x})\rangle.
\label{g1bec_lt}
\end{eqnarray}
\end{widetext}
In the second term in \eqref{g1bec_lt} we have defined
\begin{eqnarray}
\delta_{p,p^\prime}&=&
\sqrt{\frac{4\pi V}{3R_{TF}^3}}\int_{-1}^1 dz
\big|\tilde{\chi}_0\big(\sqrt{p^2+p^{\prime 2}-2pp^\prime z}\big)\big|^2\nonumber\\
&=&
\begin{cases}\Order{1} ,&|p-p^\prime|<2\pi/R_{\rm TF}\\
0,&|p-p^\prime|>2\pi/R_{\rm TF}.\end{cases}
\end{eqnarray}
As before, the contribution from the condensate is clearly dominant.
The integral in the first term in Eq. (\ref{g1bec_lt}) is
proportional to the amplitude for finding an atom pair with center
of mass momentum $\vec{p}$ and total kinetic energy $\nu$. Because
$\tilde{\chi}_0$ drops to zero on a scale of $2\pi/R_{TF}$ this
amplitude is essentially zero if $p> 2\pi/R_{\rm TF}$ or
$\nu>\frac{\pi^2}{M R_{\rm TF}^2}$.

\begin{figure}
\includegraphics{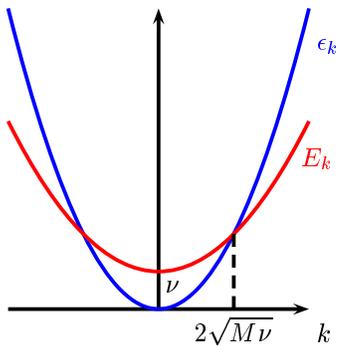}
\caption{In the narrow resonance case for momenta $p$ much larger than the
momentum width of the condensate one atom is taken out of the condensate and
the other atom is taken from the non-condensed part and has momentum close to
$p$ in order for momentum conservation to be satisfied. Since the total energy
of the atoms has to match the total energy of the molecule,  only molecules
with momenta $2\sqrt{M\nu}$ can be formed for each detuning $\nu$.}
\label{energy_conservation}
\end{figure}

\begin{figure}
\includegraphics[width=0.9\columnwidth]{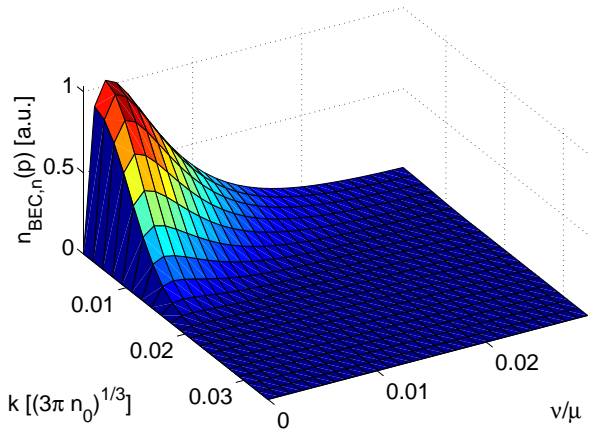}
\caption{Momentum distribution of molecules formed from a BEC of $N=10^5$ atoms
with scattering length $a=0.01a_{\rm osc}$ at $T=0.1T_c$ for a narrow
resonance.} \label{nofpBEC_lt}
\end{figure}

The second term in Eq. (\ref{g1bec_lt}) originates from molecules that
are formed from an atom in the condensate and a
non-condensed atom. Since the atom momentum $\lesssim 2\pi/R_{\rm TF}$ in the condensate
is very small 
compared to the momentum $|\vec{p}|\sim 1/\xi$ of a non-condensed atom,
the molecular momentum is essentially due to the
non-condensed atom. On the other hand, energy conservation implies
that $\nu+p^2/4M\approx p^2/2M$ if $p\gg 2\pi/R_{TF}$.
Consequently for a given detuning $\nu$, molecules with momenta in a shell
of radius $2\sqrt{M\nu}$ and width $2\pi/R_{TF}$ are formed from
one atom in the condensate and another atom taken from the
non-condensed part with a momentum that also lies in a spherical
shell in momentum space around $\vec{p}$ with thickness
$2\pi/R_{TF}$. Momentum and energy conservation are illustrated in
Fig. \ref{energy_conservation}. Figure \ref{nofpBEC_lt} shows a
typical example for the momentum distribution.

Equation \eqref{g1bec_lt} allows us to extract the criterion for the
applicability of our approximation scheme, i.e of treating the atomic state as
being undepleted. The coherent contribution will only be nonzero if
$|\vec{p}|\leq 2\pi/R_{TF}$ and for these momenta they can be neglected, as we
have seen. Requiring that the number of molecules remains much smaller then the
initial number of atoms leads to the condition
\begin{equation}
\sqrt{N}g\ll \nu \frac{1}{R_{TF}^3(M\nu)^{3/2}}.
\label{criterion_coh}
\end{equation}

In the opposite case $|\vec{p}|> 2\pi/R_{TF}$ the coherent
contribution is essentially zero and we need only consider the
incoherent contribution. Requiring that the number of molecules
with momentum $\vec{p}$ be much smaller than the number of
non-condensed atoms with that same momentum leads to
\begin{equation}
gt\ll N^{-1} \frac{\nu}{g}\frac{1}{R_{TF}^3 (M\nu)^{3/2}}.
\label{criterion_incoh}
\end{equation}

\section{\label{NFG} Normal Fermi gas}

For a normal Fermi gas we restrict ourselves to the case of zero
temperature, $T=0$. For temperatures $T$ 
well below the Fermi temperature $T_F$, the corrections to our results are of
order $(T/T_F)^2$ or higher, and do not lead to any qualitatively new effects.

Again, we treat the gas in the local density approximation where
the atoms locally fill a momentum sea
\begin{equation}
\ket{NFG}=\prod_{|\vec{k}|<k_F(\vec{x})} \hat{c}_\vec{k}^\dagger
\ket{0}
\end{equation}
with local Fermi momentum $k_F(\vec{x})$ and $\ket{0}$ being the
atomic vacuum. The Fermi momentum is related to the local chemical
potential
\begin{equation}
\mu_\text{loc}(\vec{x})=\mu_0 - U_{\rm tr}(\vec{x}) \label{local_EF}
\end{equation}
by means of
\begin{equation}
\mu_{\rm loc}(\vec{x})=\frac{k_F^2(\vec{x})}{2M}=\frac{(3\pi^2
n(\vec{x}))^{2/3}}{2M}.
\label{mulocnormal}
\end{equation}
Here, $\mu_0=(3\pi^2 n_0)^{2/3}/(2M)$ is the chemical potential
of the trapped gas, and
$n_0$ is the density at the center of the trap for each of the
spin states. The Hartree-Fock mean field has been neglected because
it gives rise only to minor corrections and doesn't lead to a qualitatively
new behavior. The density distribution of the trapped gas is given
by the Thomas-Fermi result \cite{Butts},
\begin{equation}
n(\vec{x})=\frac{N}{R_F^3}\frac{8}{\pi^2}\left[
1-\frac{r^2}{R_F^2} \right]^{3/2} \label{thomas-fermi-fermions}
\end{equation}
where $R_F=(48N)^{1/6}a_{\rm osc}$ is the Thomas-Fermi
radius for fermions.

Using the same perturbation methods as described in the
previous section for bosons, we can calculate the momentum
distribution of the molecules and their correlation function $g^{(2)}$ for a
broad and for a narrow resonance by first calculating the density of the
desired quantity at a position $\vec{x}$ and then integrating the
result over the volume of the gas.

\subsection{Broad resonance $E_{\rm kin}\ll g\sqrt{N}$}

\begin{figure}
\includegraphics{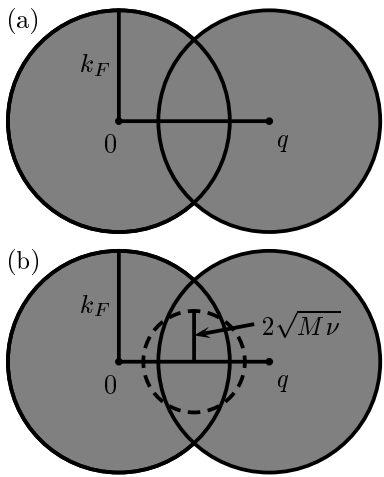}
\caption{(a) Illustration of the number of atom pairs with center
of mass momentum $q$.  These atoms can be transformed into a
molecule of momentum $q$ if energy conservation plays no role. (b)
Density of atom pairs with center of mass momentum $q$ and total
kinetic energy $2\sqrt{M\nu}$. These atom pairs can be converted
into a molecule with momentum $q$ and detuning $\nu$.} \label{F_p}
\end{figure}

To deal with the case of fermions, we modify Eq. \eqref{g1_BEC_explicit}
by reintroducing  the spin of the atoms. 
The integral over the relative momentum of
the atom pairs can be carried out exactly to give
\begin{equation}
n_{\rm NFG,b}(\vec{p},\vec{x},t)=\begin{cases}(gt)^2 F_b(\vec{p},\vec{x}),&|\vec{p}|\leq 2k_F(\vec{x})\\
0,&|\vec{p}|>2k_F(\vec{x}),
\end{cases}
\label{nofpNFG}
\end{equation}
where we have introduced the local density of atom pairs with
center of mass momentum $\vec{p}$,
\begin{equation}
F_b(\vec{p},\vec{x})=\frac{\pi k_F^3(\vec{x})}{12}
\left(16-12\frac{|\vec{p}|}{k_F(\vec{x})}+
\left(\frac{|\vec{p}|}{k_F(\vec{x})}\right)^3\right).
\end{equation}
$F_b(\vec{p},\vec{x})$ can be visualized as the integration over the
intersection of two Fermi seas shifted by $\vec{p}$ relative to
each other, as depicted in Fig. \ref{F_p}(a).

The characteristic width of the momentum distribution of the molecules is
$k_F\propto n_0^{1/3}$ which is typically much wider than the distribution
found in the BEC case. From Eq. \eqref{nofpNFG} the number of molecules
produced scales linearly with the number of atoms. This is because in contrast
to the BEC case, the molecule production is a non-collective effect.
Each atom pair is converted into a molecule independently of all
the others and there is no collective enhancement.  The integration of these
results over the volume of the cloud can easily be done numerically and is
shown in the inset in Fig. \ref{nofp_broad}.

Similarly, we can calculate the local value of $g^{(2)}$ at position $\vec{x}$, from Eq.
\eqref{general_g2},
\begin{widetext}
\begin{eqnarray}
g^{(2)}_\text{loc}(\vec{p_1}, t_1; \vec{p_2}, t_2)&=&
\label{g2p1p2fermi}
\bigg\{F_b(\vec{p_1},\vec{x})F_b(\vec{p_2},\vec{x})
-\int d\vec{k}n(\vec{p_2}/2+\vec{k}) n(\vec{p_2}/2-\vec{k})
n(\vec{p_1}-\vec{p_2}/2-\vec{k})\nonumber\\
&-&\int
d\vec{k}n(\vec{p_2}+\vec{k}-\vec{p_1}/2)
n(\vec{p_1}/2+\vec{k})n(\vec{p_1}/2-\vec{k})\\
&+&\int d\vec{k}n(\vec{p_2}-\vec{p_1}/2+\vec{k})
n(\vec{p_1}/2-\vec{k_1})n(\vec{p_2}/2-\vec{k_2}) \bigg\}\bigg/
F_b(\vec{p_1},\vec{x})F_b(\vec{p_2},\vec{x}).\nonumber
\end{eqnarray}
\end{widetext}
This result simplifies considerably for
$\vec{p}_1=\vec{p}_2=\vec{p}$,
\begin{eqnarray}
g^{(2)}_{\rm loc}(\vec{p},\vec{x},t)&\equiv&
g^{(2)}_{\rm loc}(\vec{p},t;\vec{p},t,\vec{x})\nonumber\\
&=& 2\left(1-\frac{1}{F_b(\vec{p},\vec{x})}\right).
\label{g2fermigas}
\end{eqnarray}
As in the case of the BEC, the time dependence in
$g^{(2)}(\vec{p},\vec{x},t)$ cancels at this level of
approximation. However, in contrast to the case of a BEC there is
some dependence on the momentum left.

The origin of the factor of two in $g^{(2)}(\vec{p},\vec{x},t)$ is
the following: The two molecules that are being detected in the
measurement of $g^{(2)}$ can be formed from four atoms in two
different ways and the two possibilities both give the same
contribution. Eq. (\ref{g2fermigas}) indicates that the statistics
of the molecules are super-Poissonian, similarly to a thermal field.

By the following argument we can convince ourselves that not only
the second-order correlations look thermal, but that the entire
counting statistics of each momentum mode is thermal.  Each molecular mode
characterized by the momentum $\vec{p}$ is coupled to a particular subset of
atom pairs selected by momentum conservation. In the short time limit, each
atom pair with center of mass momentum $\vec{p}$ is converted into a molecule
in the corresponding molecular mode independently of all the other atom pairs
and with uncorrelated phases. Thus we expect the number statistics of each
molecular mode to be similar to that of a light field in thermal equilibrium
with a reservoir with which there is an incoherent exchange of energy.

\subsection{Narrow resonance, $E_{kin}\gg g\sqrt{N}$}

\begin{figure}
\includegraphics[width=0.9\columnwidth]{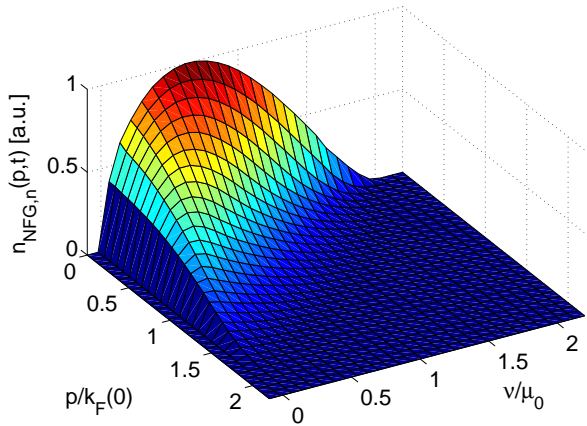}
\caption{Momentum distribution of molecules produced from a normal Fermi gas in the narrow resonance limit for $g=10^{-3}\mu$.}
\label{nofpNFG_lt}
\end{figure}

In this case, the molecules formed have to satisfy energy and momentum
conservation, as illustrated in Fig.
\ref{F_p}(b). We are then lead to a calculation
very similar to the one presented in appendix \ref{appendixb} for
the contribution from the non-condensed fraction of atoms for the
BEC case. After integrating over the volume of the cloud we find
\begin{widetext}
\begin{equation}
n_{\rm NFG,n}(\vec{p},t)=\frac{g^2t}{8\pi}M^{3/2}\nu^{1/2}\int d^3x\quad
{\rm max}\left(0,{\rm min}\left(2,\frac{k_{F}(\vec{x})^2-p^2/4-M\nu}{|\vec{p}|\sqrt{M\nu}}\right)\right).
\end{equation}
\end{widetext}
The number of molecules produced is proportional to the number of atom pairs
that satisfy momentum and energy conservation and hence scales linearly with
the number of atoms, indicating that molecule formation is not a collective
effect. Figure \ref{nofpNFG_lt} shows the momentum distribution for typical
parameters. It is much wider than the momentum distribution for the BEC case
in both momentum space and in energy width.

We don't give the lengthy and complicated expression for
$g_{\rm loc}^{(2)}(\vec{p}_1,t_1,\vec{p}_2,t_2)$ because its
qualitative properties are the same as those in the broad
resonance case except that the integration is now over pairs of
atoms that also satisfy energy conservation. For the particular
case of $\vec{p}_1=\vec{p}_2=\vec{p}$, we obtain an expression
with the exact same form as Eq. \eqref{g2fermigas} except that
$F_b(\vec{p},\vec{x})$ must be replaced with
$F_n(\vec{p},\nu,\vec{x})$,
\begin{widetext}
\begin{equation}
F_n(\vec{p},\nu,\vec{x})=
\begin{cases} \frac{M^{3/2}}{4\pi^2}\sqrt{\nu},&\sqrt{M\nu}\leq k_F(\vec{x})-p/2\\
\frac{k_{F}(\vec{x})^2-p^2/4-M\nu}{|\vec{p}|\sqrt{M\nu}},&
k_F(\vec{x})-p/2\leq \sqrt{M\nu}\leq k_F(\vec{x})+p/2\\
0,&\sqrt{M\nu}\geq k_F(\vec{x})+p/2
\end{cases}
\end{equation}
\end{widetext}
which depends on $\vec{p}$ only in the intermediate region of
detunings $k_F(\vec{x})-p/2\leq\sqrt{M\delta}\leq k_F(\vec{x})+p/2$.

\section{\label{BCS} BCS state}

Let us now consider a system of Fermions with attractive
interactions, $U_0<0$, at temperatures well below the BCS critical
temperature. As is well known, for these temperatures the
attractive interactions give rise to correlations between pairs of
atoms in time reversed states known as Cooper pairs. We assume
that the spherically symmetric trapping potential is sufficiently slowly
varying that the gas can be treated in the local density approximation. More
quantitatively, the local density approximation is valid if the size of the
Cooper pairs, given by the correlation length
\[
\lambda(r)= v_F(r)/\pi\Delta(r),
\]
is much smaller than the
oscillator length for the trap. Here, $v_F(r)$ is the velocity of
atoms at the Fermi surface and $\Delta(r)$ is the pairing field at
distance $r$ from the origin, which we take at the center of the trap.

Before turning to the coupled atom-molecule system we outline our
treatment of the atomic system. We closely follow the approach of
Houbiers et. al. \cite{Houbiers:BCS_LDA}. We assume that locally
at each $r$, the wave function can be approximated by
the BCS wave function for a homogenous gas,
\begin{equation}
\ket{BCS(r)}=\prod_{\vec{k}}(u_\vec{k}(r) +
v_\vec{k}(r)\hat{c}_{-\vec{k},\uparrow}^\dagger\hat{c}_{\vec{k},\downarrow}^\dagger)\ket{0},
\label{BCS_state}
\end{equation}
with Bogoliubov amplitudes
\begin{gather}
\label{uk}u_\vec{k}^2(r)=\frac{1}{2}\left(1+\frac{\xi_\vec{k}(r)}{\sqrt{\Delta^2(r)+\xi_\vec{k}^2(r)}}\right),\\
\label{vk}v_\vec{k}^2(r)=\frac{1}{2}\left(1-\frac{\xi_\vec{k}(r)}{\sqrt{\Delta^2(r)+\xi_\vec{k}^2(r)}}\right).
\end{gather}
Here $ \xi_\vec{k}(r)=\epsilon_\vec{k}-\mu_{\rm loc}(r)$ is the
kinetic energy of an atom measured from the local chemical
potential defined as
\begin{equation}
\mu_{loc}(r)=\mu_0-U(r)-U_0n(r).
\label{muloc}
\end{equation}
In contrast to the normal Fermi gas, we have included a Hartree-Fock
mean-field energy to the local chemical potential since we can no
longer ignore the effect of the two-body interactions in the gas.
To an excellent approximation we can use the relation Eq. \eqref{mulocnormal}
between density and local chemical potential. Then, for a given number of atoms
$N$, Eq. \eqref{muloc} is an implicit equation for $\mu_0$. We solve it
numerically and hence determine the density profile $n(r)$ and the local
chemical potential $\mu_{\rm loc}(r)$.

The gap parameter $\Delta(r)=U_0/2V\sum_k
u_{\vec{k}}(r)v_{\vec{k}}(r)$ is determined by the gap equation
\begin{equation}
\frac{-\pi}{2k_F(0) a}=\mu_0 k_F^{-3}(0)\int_0^\infty dk
k^2\left(\frac{1}{\sqrt{\xi_\vec{k}^2(r)+\Delta^2(r)}} -
\frac{1}{\xi_\vec{k}(r)}\right), \label{gapequation}
\end{equation}
where the ultra-violet divergence has been removed by
renormalizing the bare background scattering strength to the
two-body T-matrix using the Lippmann-Schwinger equation (see ref.
\cite{Houbiers:BCS_LDA}). We solve the gap equation numerically using 
the previously determined local chemical $\mu_{\rm loc}$.

\subsection{Broad resonance, $E_{kin}\ll g\sqrt{N}$}

We find the momentum distribution of the molecules from the BCS
type state by repeating the calculation done in the case of a
normal Fermi gas. For the BCS wave function, the relevant atomic
expectation values factorize as 
\begin{widetext}
\begin{eqnarray}
\left\langle \hat{c}_{\vec{p}/2-\vec{k_1},\uparrow}^\dagger
\hat{c}_{\vec{p}/2+\vec{k_1},\downarrow}^\dagger
\hat{c}_{\vec{p}/2+\vec{k_2},\uparrow}\hat{c}_{\vec{p}/2-\vec{k_2},\downarrow}
\right\rangle&=& \left\langle
\hat{c}_{\vec{p}/2-\vec{k_1},\uparrow}^\dagger
\hat{c}_{\vec{p}/2+\vec{k_1},\downarrow}^\dagger \right\rangle
\Big\langle
\hat{c}_{\vec{p}/2+\vec{k_2},\downarrow}\hat{c}_{\vec{p}/2-\vec{k_2},\uparrow}
\Big\rangle\nonumber\\
&+& \left\langle \hat{c}_{\vec{p}/2-\vec{k_1},\uparrow}^\dagger
\hat{c}_{\vec{p}/2+\vec{k_2},\downarrow} \right\rangle \Big\langle
\hat{c}_{\vec{p}/2+\vec{k_1},\downarrow}^\dagger\hat{c}_{\vec{p}/2-\vec{k_2},\uparrow}
\Big\rangle
\label{factorization_BCS}
\end{eqnarray}
\end{widetext}
and the momentum distribution of the molecules becomes
\begin{widetext}
\begin{eqnarray}
n_{\rm BCS,b}(\vec{p},t)&=& (gt)^2\left[\left|\sum_k \Big\langle
\hat{c}_{\vec{p}/2+\vec{k},\downarrow}\hat{c}_{\vec{p}/2-\vec{k},\uparrow}
\Big\rangle\right|^2 + \sum_\vec{k} \left\langle
\hat{c}_{\vec{p}/2-\vec{k},\uparrow}^\dagger
\hat{c}_{\vec{p}/2+\vec{k},\downarrow} \right\rangle \Big\langle
\hat{c}_{\vec{p}/2+\vec{k},\downarrow}^\dagger\hat{c}_{\vec{p}/2-\vec{k},\uparrow}
\Big\rangle \right]\nonumber\\
&\approx& (gt)^2\left|\sum_k \Big\langle
\hat{c}_{\vec{p}/2+\vec{k},\downarrow}\hat{c}_{\vec{p}/2-\vec{k},\uparrow}
\Big\rangle\right|^2 + n_{\rm NFG,b}(\vec{p},t). \label{nofpBCS}
\end{eqnarray}
\end{widetext}
In going from the first to the second line we have assumed that the
interactions are weak enough so that the momentum distribution of the atoms is
essentially that of a two component Fermi gas. This is justified because the
Cooper pairing only affects the momentum distribution in a small shell of
thickness $1/\lambda(r) \ll k_F(r)$ around the Fermi surface.

The first term involves the square of the pairing field. It is proportional to
the square of the number of paired atoms which, below the critical temperature,
is a finite fraction of the total number of atoms. This quadratic dependence
indicates that it is the result of a collective effect. This term can be
related to the two-point correlation function in position space as
\begin{widetext}
\begin{equation}
\Big\langle
\hat{c}_{\vec{p}/2+\vec{k},\downarrow}\hat{c}_{\vec{p}/2-\vec{k},\uparrow}
\Big\rangle =\\ \int \frac{d^3x d^3r}{V} e^{-i\vec{p}\cdot
\vec{x}-i\vec{k}\cdot\vec{r}}\langle\hat{\psi}_\downarrow(\vec{x}-\vec{r}/2)\hat{\psi}_\uparrow
(\vec{x}+\vec{r}/2)\rangle
\end{equation}
\end{widetext}
where $\hat{\psi}_{\uparrow,\downarrow}$ are the atomic field
operators in position space. In the local density approximation,
the correlation function varies with $\vec{x}$ on a length scale
$R_{\rm TF}$. On the other hand, the expectation value in the integral
falls off to zero for $r>\lambda(r)$ and we can therefore treat the
correlation function as being independent of $\vec{x}$ when
performing the integration over $\vec{r}$,
\begin{equation}
\Big\langle
\hat{c}_{\vec{p}/2+\vec{k},\downarrow}\hat{c}_{\vec{p}/2-\vec{k},\uparrow}
\Big\rangle = \int \frac{d^3x}{V} e^{-i\vec{p}\cdot\vec{x}
}\langle\hat{c}_{\vec{k},\downarrow}\hat{c}_{\vec{-k}\uparrow}\rangle\Big|_x
\end{equation}
The expectation value on the right hand side is evaluated using
the local density approximation at position $x$. Inserting the result into
Eq. \eqref{nofpBCS} and making use of the gap equation we find
\begin{widetext}
\begin{equation}
n_{\rm BCS,b}(\vec{p},t)= (gt)^2\left[\left|\int d^3x
e^{-i\vec{x}\cdot\vec{p}}
\left(1-\frac{2a\Lambda}{\pi}\right)\frac{2\Delta(x)}{U_0}\right|^2
+ n_{\rm NFG,b}(\vec{p},t)\right]. \label{BCSbroad_final}
\end{equation}
\end{widetext}
Following ref. \cite{Houbiers:BCS_LDA} we have replaced the bare
background coupling strength by
\begin{equation}
U_0=U_0\frac{1}{1-\frac{2a\Lambda}{\pi}},
\end{equation}
where $\Lambda$ is a momentum cut-off and is of the order of the
inverse of the range of the inter-atomic potential.

Using the numerically determined $\Delta(x)$ we can readily
perform the remaining Fourier transform in Eq.
\eqref{BCSbroad_final}. The result of such a calculation is shown
in Fig. \ref{nofp_broad}. Since the gap parameter changes over
distances of order $R_{TF}$ the contribution from the pairing
field has a typical width of order $1/R_{TF}$. This is very
similar to the BEC case. The background from the unpaired atoms on
the other hand has a typical width $k_F(0)=(3\pi^2 n_0)^{1/3}$
which is similar to the width of the noise contribution in the BEC
case, see inset in Fig. \ref{nofp_broad}. This similarity 
can be understood by recalling that for a weakly
interacting condensate, $n_0^{1/3}a=\beta \ll 1$, which when substituted
into the definition of the healing length gives
$1/\xi=(8\pi\beta)^{1/2}n_0^{1/3}$. For typical $\beta\sim 0.1$,
this is comparable to $k_F(0)$ for equal densities.

Because of the collective nature of the coherent contribution it
will dominate over the background, $n_{\rm NFG,b}$, for strong
enough interactions and large enough
particle numbers. The narrow width and the collective enhancement
of the molecule production are the reasons why the momentum
distribution of the molecules is such an excellent indicator of
the presence of a superfluid component and the off-diagonal long
range order accompanying it.

For weak interactions such that the coherent contribution is small compared to
the incoherent contribution, the second order correlations are close to
those of a normal Fermi gas given by Eq. \eqref{g2fermigas},
$g^{(2)}(\vec{p},\vec{x},t)\approx 2$. However, in the strongly
interacting regime, $k_F|a|\sim 1$, and large $N$, the coherent
contribution from the paired atoms dominates over the incoherent
contribution from unpaired atoms. In this limit one finds that the
second-order correlation is close to that of the BEC,
$g^{(2)}(\vec{p},\vec{x},t)\approx 1$. The physical reason for
this is that at the level of evenorder correlations
the pairing field behaves just like the mean field of
the condensate. This is clear from the factorization property of
the atomic correlation functions, Eq. \eqref{factorization_BCS},
in terms of the normal component of the density and the anomalous
density contribution due to the mean field. In this case, the
leading order terms in $N$ are given by the anomalous averages. In
the strongly interacting limit, the contribution from the
`unpaired' atoms is very similar in nature to the contribution
from the fluctuations in the BEC case.

\subsection{Narrow resonance, $E_{\rm kin}\gg g\sqrt{N}$}

\begin{figure}
\includegraphics[width=0.9\columnwidth]{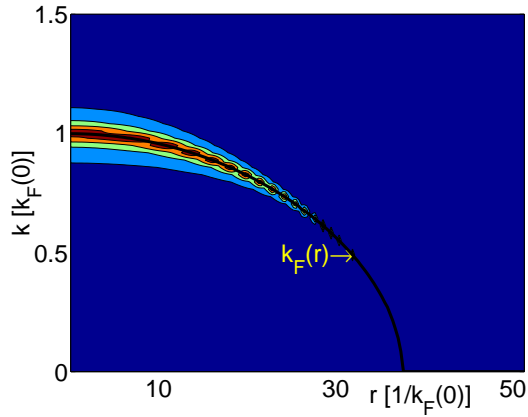}
\caption{Pairing field $\langle
\hat{c}_{\vec{k},\downarrow}\hat{c}_{-\vec{k},\uparrow}\rangle=u_k
v_k$ across the trap for $k_F a=0.5$ and $a_{\rm osc}=5k_F^{-1}(0)$.
The black solid line indicates the local Fermi momentum $k_F(r)$.}
\label{ukvkofr}
\end{figure}

A calculation similar to the one presented in Appendix
\ref{appendixb} for the BEC case leads to
\begin{widetext}
\begin{equation}
n_{\rm BCS,n}(\vec{p},t)=\left|\sum_\vec{k}\Delta(\nu
-k^2/M) \langle
\hat{c}_{\vec{p}/2+\vec{k},\downarrow}\hat{c}_{\vec{p}/2-\vec{k},\uparrow}\rangle\right|^2
+ n_{\rm NFG,n}(\vec{p},t), \label{nofpBCS_lt}
\end{equation}
\end{widetext}
where we have assumed again that the gas is weakly interacting.
Inserting Eq. \eqref{DefDelta} for $\Delta$ in the limit $\nu t\to \infty$ and
performing similar manipulations as in the broad resonance case
leads to
\begin{widetext}
\begin{eqnarray}
n_{\rm BCS,n}(\vec{p},t)&=&\frac{g^2M^3}{\pi^2p^2}
\Bigg|\int_0^\infty d\omega \sqrt{\omega}
\left(\pi\delta(\nu-\omega)+i{\bf P}\frac{1}{\nu-\omega}\right)\nonumber\\
&& \int_0^{R_{\rm TF}} dr r \sin (pr ) \langle
\hat{c}_{\sqrt{M\omega},\downarrow}\hat{c}_{-\sqrt{M\omega},\uparrow}\rangle\Big|_r\Bigg|^2
+ n_{\rm NFG,n} \label{nofpBCSnarrow_final}
\end{eqnarray}
\end{widetext}
where again the pairing field $\langle
\hat{c}_{k,\downarrow}\hat{c}_{-k,\uparrow}\rangle|_r=
u_k(r)v_k(r)$ can be evaluated using the local density
approximation. Figure \ref{ukvkofr} shows an example of the
pairing field across the trap. 
Two qualitatively different cases have to be distinguished
depending on the strength of the interactions.

\begin{figure}
\includegraphics[width=0.9\columnwidth]{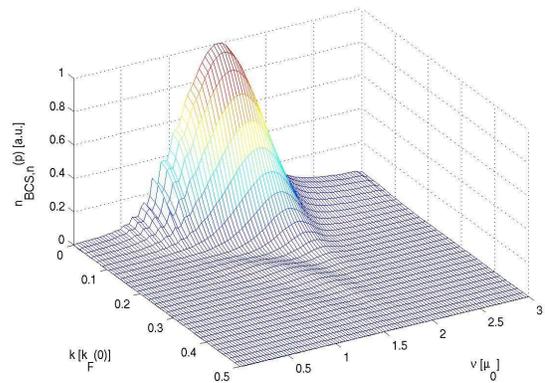}
\caption{Momentum distribution of molecules formed from a strong
coupling ($k_F a = 0.5$) BCS system in the narrow resonance regime
as a function of detuning $\nu$ for $a_{\rm osc}=5k_F^{-1}(0)$. Well below
$\nu=2\mu$ the figure is a bit noisy because the Fourier transform in eq.
\eqref{nofpBCSnarrow_final} gets its main contribution from a very narrow
region in space where the solution of the gap equation is numerically
challenging.}
\label{BCS_ns}
\end{figure}

If the interactions are fairly strong so that the pairing field,
$u_k(r) v_k(r)$, is nonzero in a rather wide region around
$k_F(r)$, the pairing field will be a slowly varying function
across the atomic cloud. Then the remaining integral in eq.
\eqref{nofpBCSnarrow_final} can be easily evaluated numerically
and we find a momentum distribution of the molecules which is
similar to the BEC case. This limit is illustrated in fig.
\ref{BCS_ns}. The width of the momentum distribution of the
molecules is again of order $1/R_{TF}$. It is known that in the strongly
interacting limit, the size of the Cooper pairs becomes comparable
to the interparticle spacing, $\lambda(r)\sim 1/k_F(r)$. The
Cooper pairs are no longer delocalized across the extent of the
cloud but now approach the limit of localized bosonic
"quasi-molecules". Thus it is not surprising that the momentum
distribution of molecules formed from a BCS type state approaches the
one we found in the BEC case.

On the other hand, if the interactions between the atoms are weak
the pairing field is a very narrow function of momentum and hence,
for fixed momentum, also of position as can be seen from Fig.
\ref{ukvkofr}. Then the integral in Eq.
\eqref{nofpBCSnarrow_final} has contributions from a rather narrow
region in space only and the momentum distribution will accordingly become wider
and smaller. 

Probing the BCS system in the narrow
resonance regime also yields spatial information about
the atomic state. By tuning $\nu=2\mu_{\rm loc}(r)$ the molecular signal is
most sensitive to the pairing field near $r$ and less sensitive to other
regions in the trap.

A qualitative difference between the BCS and BEC cases becomes
apparent if one looks at the number of molecules as a function of
the detuning. While we find that there is only a very narrow
distribution of detunings with width $\sim
\frac{(2\pi/R_{TF})^2}{2M}$ that leads to molecule formation in
the BEC case, molecules are being formed for detunings well below $\nu\sim
2\mu_0$. The
non-homogeneity of the trapped atom system manifests itself in a
completely different way in the two cases. In the BEC case, the
total energy of a particle in the condensate is just the chemical
potential while the kinetic energy of a particle is very small
compared to the mean field energy in the Thomas Fermi limit. Upon
release, the atoms in the condensate all have a spread in kinetic
energies that is of the order $\frac{(2\pi/R_{TF})^2}{2M}$ due
entirely to zero point motion. On the other hand, in the BCS case
the superfluid forms at each position near the local Fermi
momentum. Hence, atoms in the BCS state have a large energy spread
that is of order $\sim \mu_0$ with the kinetic energies of the
paired atoms being centered around $\mu_{\rm loc}(r)$ with a width $\Delta(r)$.

For the second order moment, $g^{(2)}(\vec{p},\vec{x},t)$, the
same general arguments that were put forward in the discussion of
the broad resonance also apply to the narrow resonance. In the
strongly interacting limit, the molecular field is again
approximately coherent with a noise contribution from the
unpaired fermions.

\section{\label{Summary} Summary}

We examined the momentum distribution and momentum correlations of
molecules formed by a Feshbach resonance or by photoassociation from
a quantum degenerate atomic gas. Our study elucidated the effect
of the atomic trapping potential as well as the strength of the
atom-molecule coupling relative to the characteristic energies of
the atoms on the molecular momentum distribution.

Molecules produced from an atomic BEC show a rather narrow
momentum distribution that is comparable to the zero-point
momentum width of the atomic BEC from which they are formed. In
the case of a narrow resonance, energy conservation limits the
molecules to only a narrow energy range. The molecule production
is a collective effect with contributions from all atom pairs
adding up constructively, as indicated by the quadratic scaling of
the number of molecules with the number of atoms.  Each mode of the resulting
molecular field is to a very good approximation coherent (up to terms of order
$\Order{1/N}$). The effects of noise, both due to finite temperatures and to
vacuum fluctuations, are of relative order $\Order{1/N}$. They slightly
increase the $g^{(2)}$ and cause the molecular field in each momentum state to
be only partially coherent.

In contrast, the momentum distribution of molecules formed from a
normal Fermi gas is much broader with a typical width given by the
Fermi momentum of the initial atomic cloud. The molecule
production is not collective as the number of molecules only
scales like the number of atoms rather than the square. In this
case, the second-order correlations of the molecules exhibit
super-Poissonian fluctuations, and it was argued that the molecules
are well characterized by a thermal field.

The case where molecules are produced from paired atoms in a BCS-like state
shares many properties with the BEC case: The molecule formation rate is
collective, their momentum distribution is very narrow in comparison to the
normal Fermi gas, and the molecular field is essentially coherent. The
non-collective contribution from unpaired atoms has a momentum distribution
very similar to that of the quasiparticle fluctuations in the BEC case.

In a future publication we will use the Bogoliubov-de-Gennes equations to
describe the BCS type state which is again probed with a molecular field. Thus
we will be able to go beyond the local density approximation and we can study
the BEC-BCS cross-over regime. Also, this will allow us to study the validity
of the local density approximation more carefully.

This work is supported in part by the US Office of Naval Research,
by the National Science Foundation, by the US Army Research
Office, and by the National Aeronautics and Space Administration.

\appendix

\section{\label{appendixa}Calculation of $n(\vec{p})$ for $g\gg E_{kin}$ for a BEC}

In this appendix we show how the decomposition of the atomic field operator
into condensed and non-condensed fraction together with the local density
approximation for the non-condensed part can be used to calculate the momentum
distribution of the molecules for the case of an initial BEC of atoms. More
details on the properties of fluctuations at non-zero temperatures and the use
of the local density approximation for their description can be found in
references \cite{Hutchinson:Finite_T_BEC,Reidl:Finite_T_BEC,Bergeman:BEC_Tc}.

\begin{figure}
\includegraphics{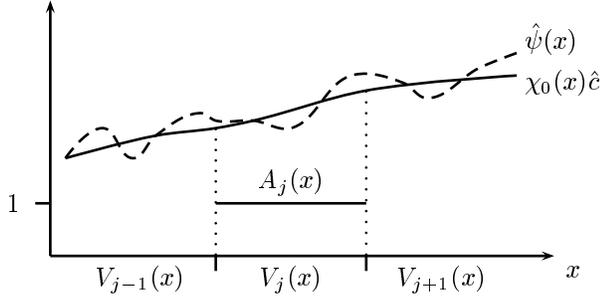}
\caption{Illustration of the partitioning of quantization volume and atomic field operator. Also indicated is the function $A_j(\vec{x})$.}
\label{partitioning_Vj}
\end{figure}

The starting point of our calculation is eq. \eqref{g1_BEC_explicit} for
the momentum distribution of the molecules. It reduces the problem to
evaluating an integral over expectation values in the undisturbed atomic field
at $t=0$. The relevant expectation values are most transparently calculated by
partitioning the quantization volume $V$ into smaller volumes $V_j$ that are
larger than the coherence length $\xi$. Partitioning the atomic field operator
accordingly, we get
\begin{equation}
\hat{\psi}(\vec{x})=\sum_j A_j(\vec{x})\chi_0(\vec{x}) \hat{c} + \sum_j
\delta\hat{\psi}_{\text{loc}}^{(j)}(\vec{x}),
\label{decomposition}
\end{equation}
where we have introduced functions $A_j(\vec{x})$ that are one in volume $V_j$
and zero otherwise, and
$\delta\hat{\psi}_{\text{loc}}^{(j)}(x)=A_j(x)\delta\hat{\psi}(x)$. The
partitioning of the field operator is illustrated in fig.
\ref{partitioning_Vj}.
We go over to the Fourier transform
\begin{equation}
\hat{c}_\vec{p}=\int\frac{d^3x}{\sqrt{V}}e^{-i\vec{p}\vec{x}}{\hat{\psi}}(\vec{x})
=\tilde{\chi}_0(\vec{p})\hat{c}+
\sum_{V_j}\sqrt{\frac{V_j}{V}}\delta\hat{c}^{(j)}_\vec{p}
\end{equation}
where $\delta\hat{c}^{(j)}(\vec{p})$ is the annihilation operator for a
fluctuation of momentum $\vec{p}$ in volume $V_j$. In the local density
approximation, fluctuations in different cells $V_j$ are uncorrelated and we
have
\begin{equation}
\langle {\delta\hat{c}_{\vec{k}_1}^{(i)}}^\dagger \delta\hat{c}_{\vec{k}_2}^{(j)}\rangle\equiv \delta_{i,j}\delta_{\vec{k}_1,\vec{k}_2}\langle {\delta\hat{c}_{\vec{k}_1}^{(i)}}^\dagger \delta\hat{c}_{\vec{k}_1}^{(i)}\rangle.
\label{uncorrelatedness}
\end{equation}
Inserting in eq.
\eqref{g1_BEC_explicit}, keeping only terms of first order in the fluctuations
and making use of relation \eqref{uncorrelatedness} we find
\begin{widetext}
\begin{multline}
n_{\rm BEC,b}(\vec{p},t)=(gt)^2 N(N-1)\sum_{\vec{k}_1,\vec{k}_2}
\tilde{\chi}_0^*(\vec{p}/2-\vec{k}_1)
\tilde{\chi}_0^*(\vec{p}/2+\vec{k}_1)
\tilde{\chi}_0(\vec{p}/2-\vec{k}_2)
\tilde{\chi}_0(\vec{p}/2+\vec{k}_2)\\
+4(gt)^2 N \sum_{\vec{k},j}\frac{V_j}{V}|\tilde{\chi}_0(\vec{p}-\vec{k})|^2
\langle \hat{c}_\vec{k}^{(j)\dagger}\hat{c}_\vec{k}^{(j)}\rangle.
\label{nofpbec_b_firststep}
\end{multline}
\end{widetext}
The coherent term is readily brought to the form given in eq. \eqref{g1bec}. In
the incoherent term we notice that $|\tilde{\chi}_0(\vec{p})|^2$ is a much
narrower function than
$\langle\hat{c}_\vec{k}^{(j)\dagger}\hat{c}_\vec{k}^{(j)}\rangle$, the former
having a typical width of $\sim 1/R_{TF}$ while the latter has a typical width
of $\sim 1/\xi$. Hence, to a good approximation,
$\langle\hat{c}_\vec{k}^{(j)\dagger}\hat{c}_\vec{k}^{(j)}\rangle$ can be
treated as a constant for the momentum range for which
$|\tilde{\chi}_0(\vec{p})|^2$ is nonzero and we obtain
\begin{multline}
4(gt)^2 N \sum_{\vec{k},j}\frac{V_j}{V}|\tilde{\chi}_0(\vec{p}-\vec{k})|^2
\langle \hat{c}_\vec{k}^{(j)\dagger}\hat{c}_\vec{k}^{(j)}\rangle\\
\approx 4(gt)^2 N \sum_{j}\frac{V_j}{V}\langle \hat{c}_\vec{k}^{(j)\dagger}\hat{c}_\vec{k}^{(j)}\rangle\sum_{\vec{k}}|\tilde{\chi}_0(\vec{p}-\vec{k})|^2\\
=4(gt)^2 N \sum_{j}\frac{V_j}{V}\langle \hat{c}_\vec{k}^{(j)\dagger}\hat{c}_\vec{k}^{(j)}\rangle,
\end{multline}
where we have made use of the normalization of $\chi_0(\vec{x})$ in the last
step. Since $\xi\ll R_{TF}$ the volumes $V_j$ can be made small and the sum
can be approximated by an integral, leading to eq. \eqref{g1bec}.

\section{\label{appendixb}Calculation of $n(\vec{p})$ for $g\ll E_{kin}$ for a BEC}

The calculation goes along similar lines as in the broad resonance case but the
evaluation of the integrals is more complicated. Repeating the calculation of
the broad resonance case that lead to eq. \eqref{nofpbec_b_firststep} with
$\hat{a}_\vec{p}$ now replaced according to eq. \eqref{aofp_lt} we find, again
to first order in the fluctuations
\begin{widetext}
\begin{multline}
n_{\rm BEC,n} (\vec{p},t)=
N(N-1)\Big|\sum_\vec{k}\Delta(E_\vec{p}-\epsilon_{\vec{p}/2+\vec{k}}-\epsilon_{\vec{p}/2-\vec{k}},t)\tilde{\chi}_0(\vec{p}/2+\vec{k})\tilde{\chi}_0(\vec{p}/2-\vec{k})\Big|^2\\
+4N\sum_{\vec{k},j}\frac{V_j}{V}\left|\Delta(E_{\vec{p}}-\epsilon_{\vec{p}/2+\vec{k}}-\epsilon_{\vec{p}/2-\vec{k}},t)\right|^2 |\tilde{\chi}_0(\vec{p}-\vec{k})|^2\langle \hat{c}_\vec{k}^{(j)\dagger}\hat{c}_\vec{k}^{(j)}\rangle\\
\equiv n_{\rm coh}(\vec{p},t) + n_{\rm incoh}(\vec{p},t).
\end{multline}
\end{widetext}
Let us first consider the coherent part $n_{\rm coh}(\vec{p},t)$.
Going over from the summation to an
integral in the usual way, making the substitution $\omega=k^2/M$ and
introducing polar coordinates we find
\begin{widetext}
\begin{multline}
n_{\rm coh}(\vec{p},t)=\frac{VM^{3/2}}{\pi^2 2^3}
\int_0^\infty d\omega \Delta(\nu-\omega,t)
\int_0^\pi d\vartheta \sin\vartheta\\
\times\sqrt{\omega}
\tilde{\chi}_0\left(\sqrt{p^2/4+M\omega+|p|\sqrt{M\omega}\cos\vartheta}\right)
\tilde{\chi}_0\left(\sqrt{p^2/4+M\omega-|p|\sqrt{M\omega}\cos\vartheta}\right)
\end{multline}
\end{widetext}

In the limit $t\rightarrow \infty$, $\Delta(\nu-\omega)$
becomes \cite{merzbacher},
\begin{equation}
\lim_{t\to \infty}\Delta(\nu-\omega,t)=g\left( \pi
\delta(\nu-\omega)-i\bf{P}\frac{1}{\nu-\omega} \right)
\label{deltafunction}
\end{equation}
where, as usual, $\bf P$ means that the integral has to be taken in the
sense of the Cauchy-principal value.
The real part can be evaluated by making use of the $\delta$-function and
for the imaginary part we have to rely on numerical methods to calculate
the principal value integral.

Making similar manipulations of the sums over momenta for the incoherent part
leads to
\begin{widetext}
\begin{multline}
n_{\rm incoh}(\vec{p},t)=4N\frac{M^{3/2}}{8\pi^2}\sum_j V_j \int_0^\pi d\vartheta \sin\vartheta\\ \times\int_0^\infty d\omega \sqrt{\omega}|\Delta(\nu-\omega,t)|^2
\Big|\tilde{\chi}_0\left(\sqrt{p^2/4+M\omega-|p|\sqrt{M\omega}\cos\vartheta}\right)\Big|^2
\langle \hat{c}_{\sqrt{M\omega}}^{(j)\dagger}\hat{c}_{\sqrt{M\omega}}^{(j)}\rangle.
\label{nincohfinal}
\end{multline}
\end{widetext}
Using the delta function
\begin{equation}
\lim_{t\to\infty}|\Delta(\nu-\omega)|^2 = \pi g^2 t\delta(\nu-\omega).
\end{equation}
to perform the integral in the limit as $t$ goes to inifinity we arrive at eq.
\eqref{g1bec_lt}.

\bibliography{draft5}

\begin{thebibliography}{42}
\expandafter\ifx\csname natexlab\endcsname\relax\def\natexlab#1{#1}\fi
\expandafter\ifx\csname bibnamefont\endcsname\relax
  \def\bibnamefont#1{#1}\fi
\expandafter\ifx\csname bibfnamefont\endcsname\relax
  \def\bibfnamefont#1{#1}\fi
\expandafter\ifx\csname citenamefont\endcsname\relax
  \def\citenamefont#1{#1}\fi
\expandafter\ifx\csname url\endcsname\relax
  \def\url#1{\texttt{#1}}\fi
\expandafter\ifx\csname urlprefix\endcsname\relax\def\urlprefix{URL }\fi
\providecommand{\bibinfo}[2]{#2}
\providecommand{\eprint}[2][]{\url{#2}}

\bibitem[{\citenamefont{Inouye et~al.}(1998)}]{Inouye:MoleculeFR}
\bibinfo{author}{\bibfnamefont{S.}~\bibnamefont{Inouye}} \bibnamefont{et~al.},
  \bibinfo{journal}{Nature (London)} \textbf{\bibinfo{volume}{392}},
  \bibinfo{pages}{151} (\bibinfo{year}{1998}).

\bibitem[{\citenamefont{Timmermans et~al.}(1999)\citenamefont{Timmermans,
  Tommasini, Hussein, and Kerman}}]{Timmermans:Feshbachresonances}
\bibinfo{author}{\bibfnamefont{E.}~\bibnamefont{Timmermans}},
  \bibinfo{author}{\bibfnamefont{P.}~\bibnamefont{Tommasini}},
  \bibinfo{author}{\bibfnamefont{M.}~\bibnamefont{Hussein}}, \bibnamefont{and}
  \bibinfo{author}{\bibfnamefont{A.}~\bibnamefont{Kerman}},
  \bibinfo{journal}{Phys. Rep.} \textbf{\bibinfo{volume}{315}},
  \bibinfo{pages}{199} (\bibinfo{year}{1999}).

\bibitem[{\citenamefont{{P. O. Fedichev} et~al.}(1996)\citenamefont{{P. O.
  Fedichev}, {Yu Kagan}, {G. V. Shlyapnikov}, and {J. T. M.
  Walraven}}}]{Fedichev:}
\bibinfo{author}{\bibnamefont{{P. O. Fedichev}}},
  \bibinfo{author}{\bibnamefont{{Yu Kagan}}}, \bibinfo{author}{\bibnamefont{{G.
  V. Shlyapnikov}}}, \bibnamefont{and} \bibinfo{author}{\bibnamefont{{J. T. M.
  Walraven}}}, \bibinfo{journal}{Phys. Rev. Lett.}
  \textbf{\bibinfo{volume}{77}}, \bibinfo{pages}{2913} (\bibinfo{year}{1996}).

\bibitem[{\citenamefont{Theis et~al.}(2004)\citenamefont{Theis, Thalhammer,
  Winkler, Hellwig, Ruff, Grimm, and {J. H. Denschlag}}}]{Theis:}
\bibinfo{author}{\bibfnamefont{M.}~\bibnamefont{Theis}},
  \bibinfo{author}{\bibfnamefont{G.}~\bibnamefont{Thalhammer}},
  \bibinfo{author}{\bibfnamefont{K.}~\bibnamefont{Winkler}},
  \bibinfo{author}{\bibfnamefont{M.}~\bibnamefont{Hellwig}},
  \bibinfo{author}{\bibfnamefont{G.}~\bibnamefont{Ruff}},
  \bibinfo{author}{\bibfnamefont{R.}~\bibnamefont{Grimm}}, \bibnamefont{and}
  \bibinfo{author}{\bibnamefont{{J. H. Denschlag}}}, \bibinfo{journal}{Phys.
  Rev. Lett.} \textbf{\bibinfo{volume}{93}}, \bibinfo{pages}{123001}
  (\bibinfo{year}{2004}).

\bibitem[{\citenamefont{Wynar et~al.}(2000)}]{Wynar:MoleculePa}
\bibinfo{author}{\bibfnamefont{R.}~\bibnamefont{Wynar}} \bibnamefont{et~al.},
  \bibinfo{journal}{Science} \textbf{\bibinfo{volume}{287}},
  \bibinfo{pages}{1016} (\bibinfo{year}{2000}).

\bibitem[{\citenamefont{Donley et~al.}(2002)\citenamefont{Donley, Claussen,
  Thompson, and Wieman}}]{Donley:MoleculeBEC}
\bibinfo{author}{\bibfnamefont{E.~A.} \bibnamefont{Donley}},
  \bibinfo{author}{\bibfnamefont{N.~R.} \bibnamefont{Claussen}},
  \bibinfo{author}{\bibfnamefont{S.~T.} \bibnamefont{Thompson}},
  \bibnamefont{and} \bibinfo{author}{\bibfnamefont{C.~E.}
  \bibnamefont{Wieman}}, \bibinfo{journal}{Nature (London)}
  \textbf{\bibinfo{volume}{417}}, \bibinfo{pages}{529} (\bibinfo{year}{2002}).

\bibitem[{\citenamefont{D\"urr et~al.}(2004)\citenamefont{D\"urr, Volz, Marte,
  and Rempe}}]{Duerr:MoleculeRP}
\bibinfo{author}{\bibfnamefont{S.}~\bibnamefont{D\"urr}},
  \bibinfo{author}{\bibfnamefont{T.}~\bibnamefont{Volz}},
  \bibinfo{author}{\bibfnamefont{A.}~\bibnamefont{Marte}}, \bibnamefont{and}
  \bibinfo{author}{\bibfnamefont{G.}~\bibnamefont{Rempe}},
  \bibinfo{journal}{Phys. Rev. Lett.} \textbf{\bibinfo{volume}{92}},
  \bibinfo{pages}{020406} (\bibinfo{year}{2004}).

\bibitem[{\citenamefont{Greiner et~al.}(2003)\citenamefont{Greiner, Regal, and
  Jin}}]{Greiner:MoleculeBEC}
\bibinfo{author}{\bibfnamefont{M.}~\bibnamefont{Greiner}},
  \bibinfo{author}{\bibfnamefont{C.~A.} \bibnamefont{Regal}}, \bibnamefont{and}
  \bibinfo{author}{\bibfnamefont{D.~S.} \bibnamefont{Jin}},
  \bibinfo{journal}{Nature (London)} \textbf{\bibinfo{volume}{426}},
  \bibinfo{pages}{537} (\bibinfo{year}{2003}).

\bibitem[{\citenamefont{Zwierlein et~al.}(2003)}]{Zwierlein:Li2}
\bibinfo{author}{\bibfnamefont{M.~W.} \bibnamefont{Zwierlein}}
  \bibnamefont{et~al.}, \bibinfo{journal}{Phys. Rev. Lett.}
  \textbf{\bibinfo{volume}{91}}, \bibinfo{pages}{250401}
  (\bibinfo{year}{2003}).

\bibitem[{\citenamefont{Jochim et~al.}(2003)}]{Jochim:Li2}
\bibinfo{author}{\bibfnamefont{S.}~\bibnamefont{Jochim}} \bibnamefont{et~al.},
  \bibinfo{journal}{Science} \textbf{\bibinfo{volume}{301}},
  \bibinfo{pages}{2101} (\bibinfo{year}{2003}).

\bibitem[{\citenamefont{Greiner
  et~al.}(2002{\natexlab{a}})\citenamefont{Greiner, Mandel, Esslinger,
  H\"ansch, and Bloch}}]{Bloch:MottInsulator1}
\bibinfo{author}{\bibfnamefont{M.}~\bibnamefont{Greiner}},
  \bibinfo{author}{\bibfnamefont{O.}~\bibnamefont{Mandel}},
  \bibinfo{author}{\bibfnamefont{T.}~\bibnamefont{Esslinger}},
  \bibinfo{author}{\bibfnamefont{T.~W.} \bibnamefont{H\"ansch}},
  \bibnamefont{and} \bibinfo{author}{\bibfnamefont{I.}~\bibnamefont{Bloch}},
  \bibinfo{journal}{Nature} \textbf{\bibinfo{volume}{415}}, \bibinfo{pages}{39}
  (\bibinfo{year}{2002}{\natexlab{a}}).

\bibitem[{\citenamefont{Greiner
  et~al.}(2002{\natexlab{b}})\citenamefont{Greiner, Mandel, H\"ansch, and
  Bloch}}]{Bloch:MottInsulator2}
\bibinfo{author}{\bibfnamefont{M.}~\bibnamefont{Greiner}},
  \bibinfo{author}{\bibfnamefont{O.}~\bibnamefont{Mandel}},
  \bibinfo{author}{\bibfnamefont{T.~W.} \bibnamefont{H\"ansch}},
  \bibnamefont{and} \bibinfo{author}{\bibfnamefont{I.}~\bibnamefont{Bloch}},
  \bibinfo{journal}{Nature} \textbf{\bibinfo{volume}{419}}, \bibinfo{pages}{51}
  (\bibinfo{year}{2002}{\natexlab{b}}).

\bibitem[{\citenamefont{Jaksch et~al.}(1998)\citenamefont{Jaksch, Bruder,
  Cirac, Gardiner, and Zoller}}]{Jaksch:BECinLattice}
\bibinfo{author}{\bibfnamefont{D.}~\bibnamefont{Jaksch}},
  \bibinfo{author}{\bibfnamefont{C.}~\bibnamefont{Bruder}},
  \bibinfo{author}{\bibfnamefont{J.~I.} \bibnamefont{Cirac}},
  \bibinfo{author}{\bibfnamefont{C.~W.} \bibnamefont{Gardiner}},
  \bibnamefont{and} \bibinfo{author}{\bibfnamefont{P.}~\bibnamefont{Zoller}},
  \bibinfo{journal}{Phys. Rev. Lett.} \textbf{\bibinfo{volume}{81}},
  \bibinfo{pages}{3108} (\bibinfo{year}{1998}).

\bibitem[{\citenamefont{Stoof}(1998)}]{Stoof:Varenna1998}
\bibinfo{author}{\bibfnamefont{H.~T.~C.} \bibnamefont{Stoof}}
  (\bibinfo{organization}{International school of physics `Enrico Fermi',
  Varenna}, \bibinfo{year}{1998}).

\bibitem[{\citenamefont{Timmermans}(2004)}]{Timmermans:Fermigasphysics}
\bibinfo{author}{\bibfnamefont{E.}~\bibnamefont{Timmermans}},
  \bibinfo{journal}{Physica Scripta} \textbf{\bibinfo{volume}{T110}},
  \bibinfo{pages}{302} (\bibinfo{year}{2004}).

\bibitem[{\citenamefont{Timmermans et~al.}(2001)\citenamefont{Timmermans,
  Furuya, Milonni, and Kerman}}]{Timmermans:BCS}
\bibinfo{author}{\bibfnamefont{E.}~\bibnamefont{Timmermans}},
  \bibinfo{author}{\bibfnamefont{K.}~\bibnamefont{Furuya}},
  \bibinfo{author}{\bibfnamefont{P.~W.} \bibnamefont{Milonni}},
  \bibnamefont{and} \bibinfo{author}{\bibfnamefont{A.~K.}
  \bibnamefont{Kerman}}, \bibinfo{journal}{Phys. Lett. A}
  \textbf{\bibinfo{volume}{285}}, \bibinfo{pages}{288} (\bibinfo{year}{2001}).

\bibitem[{\citenamefont{Regal et~al.}(2004)\citenamefont{Regal, Greiner, and
  Jin}}]{Regal:BEC_BCScrossover}
\bibinfo{author}{\bibfnamefont{C.~A.} \bibnamefont{Regal}},
  \bibinfo{author}{\bibfnamefont{M.}~\bibnamefont{Greiner}}, \bibnamefont{and}
  \bibinfo{author}{\bibfnamefont{D.~S.} \bibnamefont{Jin}},
  \bibinfo{journal}{Phys. Rev. Lett.} \textbf{\bibinfo{volume}{92}},
  \bibinfo{pages}{040403} (\bibinfo{year}{2004}).

\bibitem[{\citenamefont{Bartenstein
  et~al.}(2004{\natexlab{a}})}]{Bartenstein:BEC_BCScrossover}
\bibinfo{author}{\bibfnamefont{M.}~\bibnamefont{Bartenstein}}
  \bibnamefont{et~al.}, \bibinfo{journal}{Phys. Rev. Lett.}
  \textbf{\bibinfo{volume}{92}}, \bibinfo{pages}{120401}
  (\bibinfo{year}{2004}{\natexlab{a}}).

\bibitem[{\citenamefont{Zwierlein et~al.}(2004)}]{Zwierlein:BEC_BCScrossover}
\bibinfo{author}{\bibfnamefont{M.}~\bibnamefont{Zwierlein}}
  \bibnamefont{et~al.}, \bibinfo{journal}{Phys. Rev. Lett.}
  \textbf{\bibinfo{volume}{92}}, \bibinfo{pages}{120403}
  (\bibinfo{year}{2004}).

\bibitem[{\citenamefont{Stoof et~al.}(1996)\citenamefont{Stoof, Houbiers,
  Sackett, and Hulet}}]{Stoof:Li6}
\bibinfo{author}{\bibfnamefont{H.~T.~C.} \bibnamefont{Stoof}},
  \bibinfo{author}{\bibfnamefont{M.}~\bibnamefont{Houbiers}},
  \bibinfo{author}{\bibfnamefont{C.~A.} \bibnamefont{Sackett}},
  \bibnamefont{and} \bibinfo{author}{\bibfnamefont{R.~G.} \bibnamefont{Hulet}},
  \bibinfo{journal}{Phys. Rev. Lett.} \textbf{\bibinfo{volume}{76}},
  \bibinfo{pages}{10} (\bibinfo{year}{1996}).

\bibitem[{\citenamefont{Bruun et~al.}(1999)\citenamefont{Bruun, Castin, Dum,
  and Burnett}}]{Bruun:BCS_theory}
\bibinfo{author}{\bibfnamefont{G.}~\bibnamefont{Bruun}},
  \bibinfo{author}{\bibfnamefont{Y.}~\bibnamefont{Castin}},
  \bibinfo{author}{\bibfnamefont{R.}~\bibnamefont{Dum}}, \bibnamefont{and}
  \bibinfo{author}{\bibfnamefont{K.}~\bibnamefont{Burnett}},
  \bibinfo{journal}{Eur. Phys. J. D} \textbf{\bibinfo{volume}{7}},
  \bibinfo{pages}{433} (\bibinfo{year}{1999}).

\bibitem[{\citenamefont{Chin et~al.}(2004)\citenamefont{Chin, Bartenstein,
  Altmeyer, Riedl, Jochim, {J. Hecker Denschlag}, and Grimm}}]{Chin:}
\bibinfo{author}{\bibfnamefont{C.}~\bibnamefont{Chin}},
  \bibinfo{author}{\bibfnamefont{M.}~\bibnamefont{Bartenstein}},
  \bibinfo{author}{\bibfnamefont{A.}~\bibnamefont{Altmeyer}},
  \bibinfo{author}{\bibfnamefont{S.}~\bibnamefont{Riedl}},
  \bibinfo{author}{\bibfnamefont{S.}~\bibnamefont{Jochim}},
  \bibinfo{author}{\bibnamefont{{J. Hecker Denschlag}}}, \bibnamefont{and}
  \bibinfo{author}{\bibfnamefont{R.}~\bibnamefont{Grimm}},
  \bibinfo{journal}{Science} \textbf{\bibinfo{volume}{305}},
  \bibinfo{pages}{1128} (\bibinfo{year}{2004}).

\bibitem[{\citenamefont{Kinast et~al.}(2004)\citenamefont{Kinast, {S. L.
  Hemmer}, {M. E. Gehm}, Turlapov, and {J. E. Thomas}}}]{Kinast:}
\bibinfo{author}{\bibfnamefont{J.}~\bibnamefont{Kinast}},
  \bibinfo{author}{\bibnamefont{{S. L. Hemmer}}},
  \bibinfo{author}{\bibnamefont{{M. E. Gehm}}},
  \bibinfo{author}{\bibfnamefont{A.}~\bibnamefont{Turlapov}}, \bibnamefont{and}
  \bibinfo{author}{\bibnamefont{{J. E. Thomas}}}, \bibinfo{journal}{Phys. Rev.
  Lett.} \textbf{\bibinfo{volume}{92}}, \bibinfo{pages}{150402}
  (\bibinfo{year}{2004}).

\bibitem[{\citenamefont{Bartenstein
  et~al.}(2004{\natexlab{b}})\citenamefont{Bartenstein, Altmeyer, Riedl,
  Jochim, Chin, {J. H. Denschlag}, and Grimm}}]{Bartenstein:}
\bibinfo{author}{\bibfnamefont{M.}~\bibnamefont{Bartenstein}},
  \bibinfo{author}{\bibfnamefont{A.}~\bibnamefont{Altmeyer}},
  \bibinfo{author}{\bibfnamefont{S.}~\bibnamefont{Riedl}},
  \bibinfo{author}{\bibfnamefont{S.}~\bibnamefont{Jochim}},
  \bibinfo{author}{\bibfnamefont{C.}~\bibnamefont{Chin}},
  \bibinfo{author}{\bibnamefont{{J. H. Denschlag}}}, \bibnamefont{and}
  \bibinfo{author}{\bibfnamefont{R.}~\bibnamefont{Grimm}},
  \bibinfo{journal}{Phys. Rev. Lett.} \textbf{\bibinfo{volume}{92}},
  \bibinfo{pages}{203201} (\bibinfo{year}{2004}{\natexlab{b}}).

\bibitem[{\citenamefont{Burt et~al.}(1997)\citenamefont{Burt, Ghrist, Myatt,
  Holland, Cornell, and Wieman}}]{Burt:BEC_decay}
\bibinfo{author}{\bibfnamefont{E.~A.} \bibnamefont{Burt}},
  \bibinfo{author}{\bibfnamefont{R.~W.} \bibnamefont{Ghrist}},
  \bibinfo{author}{\bibfnamefont{C.~J.} \bibnamefont{Myatt}},
  \bibinfo{author}{\bibfnamefont{M.~J.} \bibnamefont{Holland}},
  \bibinfo{author}{\bibfnamefont{E.~A.} \bibnamefont{Cornell}},
  \bibnamefont{and} \bibinfo{author}{\bibfnamefont{C.~E.}
  \bibnamefont{Wieman}}, \bibinfo{journal}{Phys. Rev. Lett.}
  \textbf{\bibinfo{volume}{79}}, \bibinfo{pages}{337} (\bibinfo{year}{1997}).

\bibitem[{\citenamefont{Hellweg et~al.}(2003)\citenamefont{Hellweg,
  Cacciapuoti, Kottke, Schulte, Sengstock, Ertmer, and
  Arlt}}]{Hellweg:Spatial_correlations}
\bibinfo{author}{\bibfnamefont{D.}~\bibnamefont{Hellweg}},
  \bibinfo{author}{\bibfnamefont{L.}~\bibnamefont{Cacciapuoti}},
  \bibinfo{author}{\bibfnamefont{M.}~\bibnamefont{Kottke}},
  \bibinfo{author}{\bibfnamefont{T.}~\bibnamefont{Schulte}},
  \bibinfo{author}{\bibfnamefont{K.}~\bibnamefont{Sengstock}},
  \bibinfo{author}{\bibfnamefont{W.}~\bibnamefont{Ertmer}}, \bibnamefont{and}
  \bibinfo{author}{\bibfnamefont{J.~J.} \bibnamefont{Arlt}},
  \bibinfo{journal}{Phys. Rev. Lett.} \textbf{\bibinfo{volume}{91}},
  \bibinfo{pages}{010406} (\bibinfo{year}{2003}).

\bibitem[{\citenamefont{Cacciapuoti et~al.}(2003)\citenamefont{Cacciapuoti,
  Hellweg, Kottke, Schulte, Ertmer, Arlt, Sengstock, and
  Santos}}]{Cacciapuoti:Second_correlation_function}
\bibinfo{author}{\bibfnamefont{L.}~\bibnamefont{Cacciapuoti}},
  \bibinfo{author}{\bibfnamefont{D.}~\bibnamefont{Hellweg}},
  \bibinfo{author}{\bibfnamefont{M.}~\bibnamefont{Kottke}},
  \bibinfo{author}{\bibfnamefont{T.}~\bibnamefont{Schulte}},
  \bibinfo{author}{\bibfnamefont{W.}~\bibnamefont{Ertmer}},
  \bibinfo{author}{\bibfnamefont{J.~J.} \bibnamefont{Arlt}},
  \bibinfo{author}{\bibfnamefont{K.}~\bibnamefont{Sengstock}},
  \bibnamefont{and} \bibinfo{author}{\bibfnamefont{L.}~\bibnamefont{Santos}},
  \bibinfo{journal}{Phys. Rev. A} \textbf{\bibinfo{volume}{68}},
  \bibinfo{pages}{053612} (\bibinfo{year}{2003}).

\bibitem[{\citenamefont{Altman et~al.}(2004)\citenamefont{Altman, Demler, and
  Lukin}}]{Altman:noise_correlations}
\bibinfo{author}{\bibfnamefont{E.}~\bibnamefont{Altman}},
  \bibinfo{author}{\bibfnamefont{E.}~\bibnamefont{Demler}}, \bibnamefont{and}
  \bibinfo{author}{\bibfnamefont{M.~D.} \bibnamefont{Lukin}},
  \bibinfo{journal}{Phys. Rev. A} \textbf{\bibinfo{volume}{70}},
  \bibinfo{pages}{013603} (\bibinfo{year}{2004}).

\bibitem[{\citenamefont{Bach and
  Rza\.{z}ewski}(2004)}]{Radka:Diagnosing_correlations}
\bibinfo{author}{\bibfnamefont{R.}~\bibnamefont{Bach}} \bibnamefont{and}
  \bibinfo{author}{\bibfnamefont{K.}~\bibnamefont{Rza\.{z}ewski}},
  \bibinfo{journal}{Phys. Rev. Lett.} \textbf{\bibinfo{volume}{92}},
  \bibinfo{pages}{200401} (\bibinfo{year}{2004}).

\bibitem[{\citenamefont{Glauber}(1963{\natexlab{a}})}]{Glauber:Optical_Coheren%
ce1}
\bibinfo{author}{\bibfnamefont{R.~J.} \bibnamefont{Glauber}},
  \bibinfo{journal}{Phys. Rev.} \textbf{\bibinfo{volume}{130}},
  \bibinfo{pages}{2529} (\bibinfo{year}{1963}{\natexlab{a}}).

\bibitem[{\citenamefont{Glauber}(1963{\natexlab{b}})}]{Glauber:Optical_Coheren%
ce2}
\bibinfo{author}{\bibfnamefont{R.~J.} \bibnamefont{Glauber}},
  \bibinfo{journal}{Phys. Rev.} \textbf{\bibinfo{volume}{131}},
  \bibinfo{pages}{2766} (\bibinfo{year}{1963}{\natexlab{b}}).

\bibitem[{\citenamefont{Naraschewski and
  Glauber}(1999)}]{Naraschewski:Spatial_coherence}
\bibinfo{author}{\bibfnamefont{M.}~\bibnamefont{Naraschewski}}
  \bibnamefont{and} \bibinfo{author}{\bibfnamefont{R.~J.}
  \bibnamefont{Glauber}}, \bibinfo{journal}{Phys. Rev. A}
  \textbf{\bibinfo{volume}{59}}, \bibinfo{pages}{4595} (\bibinfo{year}{1999}).

\bibitem[{\citenamefont{Cahill and
  Glauber}(1999)}]{Cahill:Density_Operator_Fermions}
\bibinfo{author}{\bibfnamefont{K.~E.} \bibnamefont{Cahill}} \bibnamefont{and}
  \bibinfo{author}{\bibfnamefont{R.~J.} \bibnamefont{Glauber}},
  \bibinfo{journal}{Phys. Rev. A} \textbf{\bibinfo{volume}{59}},
  \bibinfo{pages}{1538} (\bibinfo{year}{1999}).

\bibitem[{\citenamefont{Meiser and
  Meystre}(2004)}]{Meiser:singlemode_molecules}
\bibinfo{author}{\bibfnamefont{D.}~\bibnamefont{Meiser}} \bibnamefont{and}
  \bibinfo{author}{\bibfnamefont{P.}~\bibnamefont{Meystre}}
  (\bibinfo{year}{2004}), \eprint{cond-mat/0410349}.

\bibitem[{\citenamefont{Holland et~al.}(2001)\citenamefont{Holland, Kokkelmans,
  Chiofalo, and Walser}}]{Holland:BCS}
\bibinfo{author}{\bibfnamefont{M.}~\bibnamefont{Holland}},
  \bibinfo{author}{\bibfnamefont{S.~J. J. M.~F.} \bibnamefont{Kokkelmans}},
  \bibinfo{author}{\bibfnamefont{M.~L.} \bibnamefont{Chiofalo}},
  \bibnamefont{and} \bibinfo{author}{\bibfnamefont{R.}~\bibnamefont{Walser}},
  \bibinfo{journal}{Phys. Rev. Lett.} \textbf{\bibinfo{volume}{87}},
  \bibinfo{pages}{120406} (\bibinfo{year}{2001}).

\bibitem[{\citenamefont{Chiofalo et~al.}(2002)\citenamefont{Chiofalo,
  Kokkelmans, Milstein, and Holland}}]{Chiofalo:Res_superfluidity}
\bibinfo{author}{\bibfnamefont{M.~L.} \bibnamefont{Chiofalo}},
  \bibinfo{author}{\bibfnamefont{S.~J. J. M.~F.} \bibnamefont{Kokkelmans}},
  \bibinfo{author}{\bibfnamefont{J.~N.} \bibnamefont{Milstein}},
  \bibnamefont{and} \bibinfo{author}{\bibfnamefont{M.~J.}
  \bibnamefont{Holland}}, \bibinfo{journal}{Phys. Rev. Lett.}
  \textbf{\bibinfo{volume}{88}}, \bibinfo{pages}{090402}
  (\bibinfo{year}{2002}).

\bibitem[{\citenamefont{{D. A. W. Hutchinson} et~al.}(1997)\citenamefont{{D. A.
  W. Hutchinson}, Zaremba, and Griffin}}]{Hutchinson:Finite_T_BEC}
\bibinfo{author}{\bibnamefont{{D. A. W. Hutchinson}}},
  \bibinfo{author}{\bibfnamefont{E.}~\bibnamefont{Zaremba}}, \bibnamefont{and}
  \bibinfo{author}{\bibfnamefont{A.}~\bibnamefont{Griffin}},
  \bibinfo{journal}{Phys. Rev. Lett.} \textbf{\bibinfo{volume}{78}},
  \bibinfo{pages}{1842} (\bibinfo{year}{1997}).

\bibitem[{\citenamefont{Reidl et~al.}(1999)\citenamefont{Reidl, Csord\'as,
  Graham, and Sz\'epfalusy}}]{Reidl:Finite_T_BEC}
\bibinfo{author}{\bibfnamefont{J.}~\bibnamefont{Reidl}},
  \bibinfo{author}{\bibfnamefont{A.}~\bibnamefont{Csord\'as}},
  \bibinfo{author}{\bibfnamefont{R.}~\bibnamefont{Graham}}, \bibnamefont{and}
  \bibinfo{author}{\bibfnamefont{P.}~\bibnamefont{Sz\'epfalusy}},
  \bibinfo{journal}{Phys. Rev. A} \textbf{\bibinfo{volume}{59}},
  \bibinfo{pages}{3816} (\bibinfo{year}{1999}).

\bibitem[{\citenamefont{Bergeman et~al.}(2000)\citenamefont{Bergeman, Feder,
  Balazs, and Schneider}}]{Bergeman:BEC_Tc}
\bibinfo{author}{\bibfnamefont{T.}~\bibnamefont{Bergeman}},
  \bibinfo{author}{\bibfnamefont{D.~L.} \bibnamefont{Feder}},
  \bibinfo{author}{\bibfnamefont{N.~L.} \bibnamefont{Balazs}},
  \bibnamefont{and} \bibinfo{author}{\bibfnamefont{B.~I.}
  \bibnamefont{Schneider}}, \bibinfo{journal}{Phys. Rev. A}
  \textbf{\bibinfo{volume}{61}}, \bibinfo{pages}{063605}
  (\bibinfo{year}{2000}).

\bibitem[{\citenamefont{{D. A. Butts} and {D. S. Rokhsar}}(1997)}]{Butts}
\bibinfo{author}{\bibnamefont{{D. A. Butts}}} \bibnamefont{and}
  \bibinfo{author}{\bibnamefont{{D. S. Rokhsar}}}, \bibinfo{journal}{Phys. Rev.
  A} \textbf{\bibinfo{volume}{55}}, \bibinfo{pages}{4346}
  (\bibinfo{year}{1997}).

\bibitem[{\citenamefont{Houbiers et~al.}(1997)}]{Houbiers:BCS_LDA}
\bibinfo{author}{\bibfnamefont{M.}~\bibnamefont{Houbiers}}
  \bibnamefont{et~al.}, \bibinfo{journal}{Phys. Rev. A}
  \textbf{\bibinfo{volume}{56}}, \bibinfo{pages}{4864} (\bibinfo{year}{1997}).

\bibitem[{\citenamefont{Merzbacher}(1998)}]{merzbacher}
\bibinfo{author}{\bibfnamefont{E.}~\bibnamefont{Merzbacher}},
  \emph{\bibinfo{title}{Quantum Mechanics}} (\bibinfo{publisher}{John Wiley \&
  Sons, New York}, \bibinfo{year}{1998}), \bibinfo{edition}{3rd} ed.

\end{thebibliography}

\end{document}